\RequirePackage{fix-cm}
\documentclass[twocolumn,english]{template/svjour3}          
\usepackage{graphicx}
\usepackage{xspace}
\usepackage[english]{babel}
\usepackage[sort&compress,numbers]{natbib} 
\usepackage{amsmath,amssymb, revsymb}
\usepackage{physics}
\usepackage{color}
\usepackage{isotope}
\usepackage{hyperref}
\definecolor{linkcolor}{rgb}{0,0,1.00} 
\hypersetup{%
    pdfsubject=Paper,
    pdfkeywords={nuclear physics} {Bayesian} {chiral EFT} {emulator},
    unicode = true,
    breaklinks = true,
    colorlinks = true,
    linkcolor = linkcolor,
    citecolor = linkcolor,
    menucolor = linkcolor,
    urlcolor = linkcolor
}

\graphicspath{{./figures/}}

\makeatletter
\newcommand\newsubcommand[3]{\newcommand#1{#2\sc@sub{#3}}}
\def\sc@sub#1{\def\sc@thesub{#1}\@ifnextchar_{\sc@mergesubs}{_{\sc@thesub}}}
\def\sc@mergesubs_#1{_{\sc@thesub#1}}

\newcommand\newsupcommand[3]{\newcommand#1{#2\sc@sup{#3}}}
\def\sc@sup#1{\def\sc@thesup{#1}\@ifnextchar^{\sc@mergesups}{^{\sc@thesup}}}
\def\sc@mergesups^#1{^{\sc@thesup#1}}
\makeatother

\DeclareMathAlphabet{\mathbcal}{OMS}{cmsy}{b}{n}

\newcommand{\etal}{\textit{et~al.}\xspace}

\newcommand{\Abinitio}{\textit{Ab~initio}\xspace}
\newcommand{\abinitio}{\textit{ab~initio}\xspace}

\newcommand{\fmiq}{\, \text{fm}^{-3}}

\newcommand{\MeV}{\, \text{MeV}}

\newcommand{\NkLO}[1]{\ensuremath{\mathrm{N}^{#1}\mathrm{LO}}\xspace}

\newcommand{\ordervec}{\vec}

\newcommand{\inputvec}{\mathbf}

\newsubcommand{\ckvec}{\ordervec{c}}{k}

\newsubcommand{\bkvec}{\ordervec{b}}{k}

\newsubcommand{\ckvecset}{\ordervec{\inputvec{c}}}{k}

\newsubcommand{\ckvecapprox}{\mathbf{c}'}{k}
\newsubcommand{\ckvecapproxset}{\mathbf{C}'}{k}

\newsubcommand{\bkvecapprox}{\mathbf{b}'}{k}
\newsubcommand{\bkvecset}{\mathbf{B}}{k}
\newsubcommand{\bkvecapproxset}{\mathbf{B}'}{k}

\newcommand{\genobs}{y}

\newsubcommand{\genobsvec}{\ordervec{\genobs}}{k}
\newsubcommand{\genobsvecset}{\ordervec{\inputvec{\genobs}}}{k}

\newsubcommand{\akvec}{\mathbf{a}}{k}

\newsubcommand{\akvecapprox}{\mathbf{a}'}{k}
\newsubcommand{\akvecset}{\mathbf{A}}{k}
\newsubcommand{\akvecapproxset}{\mathbf{A}'}{k}

{}  

\def\diffd{\mathrm{d}}  

\DeclareDocumentCommand\differential{ o g d() }{ 
    \IfNoValueTF{#2}{
        \IfNoValueTF{#3}
            {\diffd\IfNoValueTF{#1}{}{^{#1}}}
            {\mathinner{\diffd\IfNoValueTF{#1}{}{^{#1}}\argopen(#3\argclose)}}
        }
        {\mathinner{\diffd\IfNoValueTF{#1}{}{^{#1}}#2} \IfNoValueTF{#3}{}{(#3)}}
    }

\newcommand{\pathd}{\mathcal{D}}  

\DeclareDocumentCommand\pathdifferential{ o g d() }{ 
    \IfNoValueTF{#2}{
        \IfNoValueTF{#3}
            {\pathd\IfNoValueTF{#1}{}{^{#1}}}
            {\mathinner{\pathd\IfNoValueTF{#1}{}{^{#1}}\argopen(#3\argclose)}}
        }
        {\mathinner{\pathd\IfNoValueTF{#1}{}{^{#1}}#2} \IfNoValueTF{#3}{}{(#3)}}
    }

\newcommand{\eg}{\textit{e.g.}}
\newcommand{\ie}{\textit{i.e.}}
\newcommand{\nsat}{n_\text{sat}}

\begin{document}

\title{A brief account of Steven Weinberg's legacy in \abinitio many-body theory
}
\subtitle{Special issue in Few-Body Systems: Celebrating 30 years of Steven Weinberg's papers on Nuclear Forces from Chiral Lagrangians}


\author{Christian Drischler \and Scott K. Bogner}


\institute{Christian Drischler \at
              Facility for Rare Isotope Beams\\
              Michigan State University\\
              East Lansing, MI~48824, USA\\
              \email{\href{mailto:drischler@frib.msu.edu}{drischler@frib.msu.edu}}  \and
           Scott K. Bogner \at
              Facility for Rare Isotope Beams and\\
              Department of Physics and Astronomy\\
              Michigan State University\\
              East Lansing, MI~48824, USA\\
              \email{\href{mailto:bogner@frib.msu.edu}{bogner@frib.msu.edu}} 
}

\date{Received: August 8, 2021~/~Accepted: August 27, 2021}

\maketitle

\begin{abstract}

In this contribution to the special issue ``Celebrating 30 years of Steven Weinberg's papers on Nuclear Forces from Chiral Lagrangians,'' we  emphasize the important role chiral effective field theory has played in leading nuclear physics into a precision era. To this end, we share our perspective on a few of the recent advances made in \abinitio calculations of nuclear structure and nuclear matter observables, as well as Bayesian uncertainty quantification of effective field theory truncation errors.


\end{abstract}

\section{Introduction} \label{sec:intro}

On July~23, 2021, by the time we were completing this manuscript, Nobel laureate Steven Weinberg died at the age of 88~\cite{WeinbergUTAus}.
With his passing the physics community has lost one of its greatest minds, a co-founder of the Standard Model, and the father of nuclear Effective Field Theory (EFT)---in particular, chiral EFT, for which his seminal papers~\cite{Weinberg:1990rz,Weinberg:1991um,Weinberg:1992yk} in the early 1990s laid the groundwork.

\begin{figure}[tb]
	\begin{centering}
		\includegraphics[width=\linewidth]{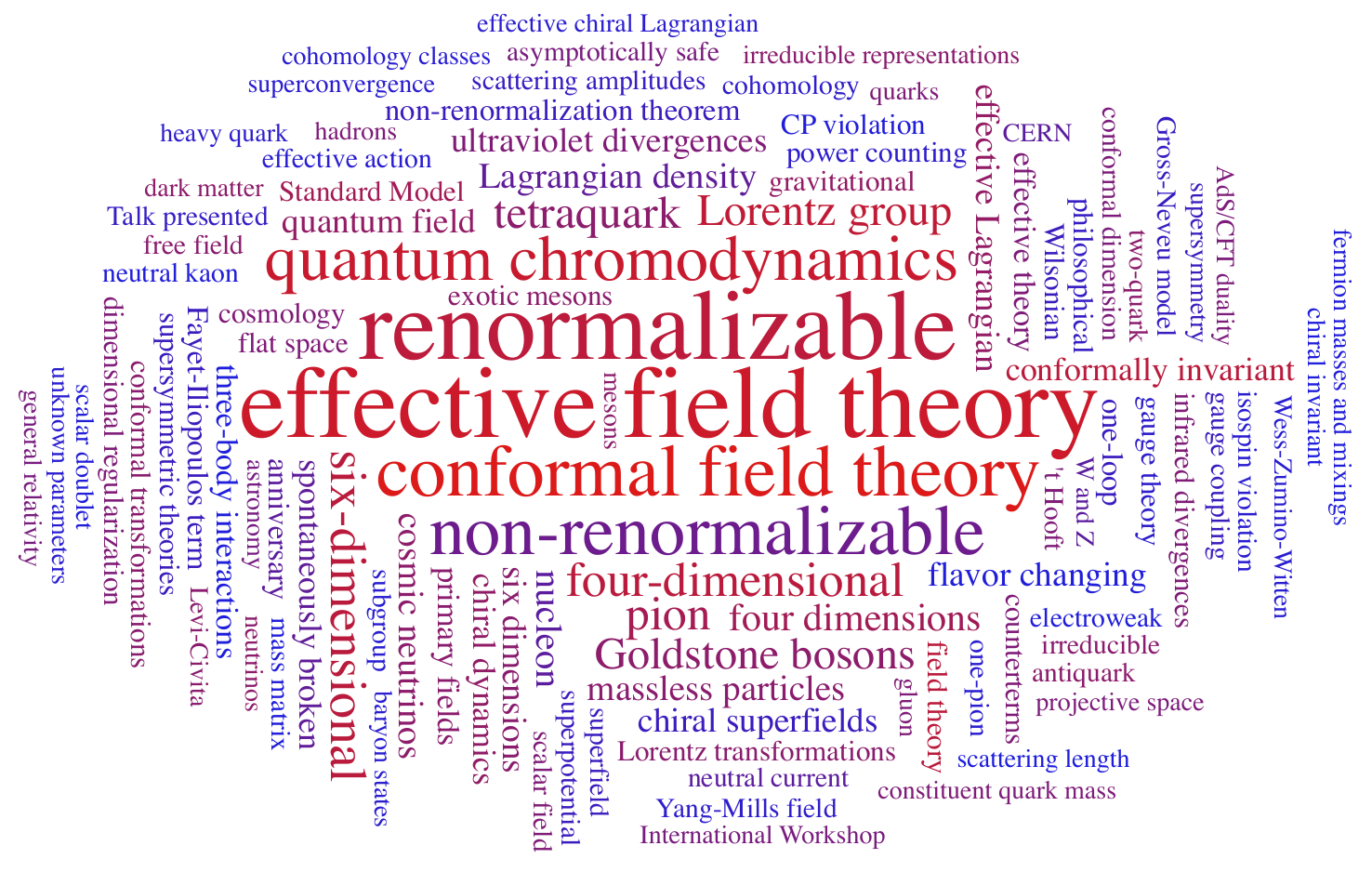}
	\end{centering}
	\caption{
	 Keyword cloud automatically generated using the tool \textsc{Scimeter}~\cite{Scimeterweb} and Weinberg's papers on the arXiv preprint server.
	 His important contributions to other fields, such as cosmology, have been suppressed in favor of the subject of this special issue.
	 Weinberg's first paper posted on the arXiv was the 1992 seminal paper on three-body interactions among nucleons and pions~\cite{Weinberg:1992yk}.
	}
	\label{fig:word_cloud}
\end{figure}

In this contribution to the special issue ``Celebrating 30 years of Steven Weinberg's papers on Nuclear Forces from Chiral Lagrangians,'' we honor Weinberg's scientific achievements by emphasizing the important role chiral EFT has played in leading nuclear physics into a precision era~\cite{Epelbaum:2015pfa}.
But where to begin?
And where to end---given the wide range of keywords in Figure~\ref{fig:word_cloud} that is associated with Weinberg's legacy? 
To this end, we share here our personal view on only a few recent advances made in \abinitio calculations of nuclear structure and nuclear matter, which have been fueled by several key advantages that chiral EFT offers over older phenomenological nuclear approaches:
\begin{itemize}
    \item Nuclear forces and currents can be consistently derived based on the symmetries of low-energy quantum chromodynamics (QCD) and then used to make predictions for nuclear observables from first principles (\ie, \abinitio calculations).
    Three-nucleon (3N) forces appear naturally and at a higher order in the chiral expansion than the leading nucleon-nucleon (NN) forces.
    
    \item Chiral NN and 3N potentials are relatively soft, which leads to important computational simplifications in most many-body calculations. The hard core (\ie, strong short-range repulsion) typically found in phenomenological potentials is suppressed in chiral potentials by regulator functions with moderate momentum cutoffs.

    \item Theoretical uncertainties can be rigorously quantified by order-by-order calculations combined with Bayesian statistical methods. Chiral EFT predicts the momentum scale at which it breaks down.
\end{itemize}

This article is organized as follows. In Sections~\ref{sec:workflow} to~\ref{sec:uncertainties} we will point out the importance of each of those (interrelated) advantages using recent many-body applications as examples. We also discuss promising new tools that will shed light on some of the issues inherent in chiral EFT with Weinberg power counting. 
The article concludes with a brief summary in Section~\ref{sec:summary}.

For more details on the broad subjects covered in this article, we share here our (incomplete) personal hot list of interesting articles with the reader: 
\begin{itemize}
    \item (chiral) EFT in general~\cite{Epelbaum:2019kcf,Hammer:2019poc,Machleidt:2011zz,Epelbaum:2008ga}
    
    \item implementation and importance of chiral 3N interactions~\cite{Hebeler:2020ocj,Hebeler:2015hla}
    
    \item Renormalization Group methods in low-energy nuclear physics~\cite{Bogner:2009bt,Furnstahl:2013vda}
    
    \item nuclear equation of state and astrophysical applications~\cite{Lattimer2021annrev,Drischler:2021kxf,Tews:2020wrl}
    
    \item current status of \abinitio nuclear structure calculations~\cite{Hergert:2020bxy,Stroberg:2019mxo}
    
    \item Bayesian methods~\cite{Sivia:2006} for uncertainty quantification~\cite{Melendez:2020xcs,Phillips:2020dmw,Melendez:2019izc,Furnstahl:2014xsa}
    
    \item eigenvector continuation~\cite{Sarkar:2021fpz,Frame:2017fah} and applications of emulators~\cite{Melendez:2021lyq,Wesolowski:2021cni,Miller:2021pcu,Furnstahl:2020abp,Ekstrom:2019lss,Konig:2019adq}
    
    \item applications of delta-full chiral EFT~\cite{Jiang:2020the,Piarulli:2019cqu,Ekstrom:2017koy}
\end{itemize}

%
\section{\Abinitio workflow in many-body theory} \label{sec:workflow}

\begin{figure}[tb]
	\begin{center}
	    \includegraphics[scale=1]{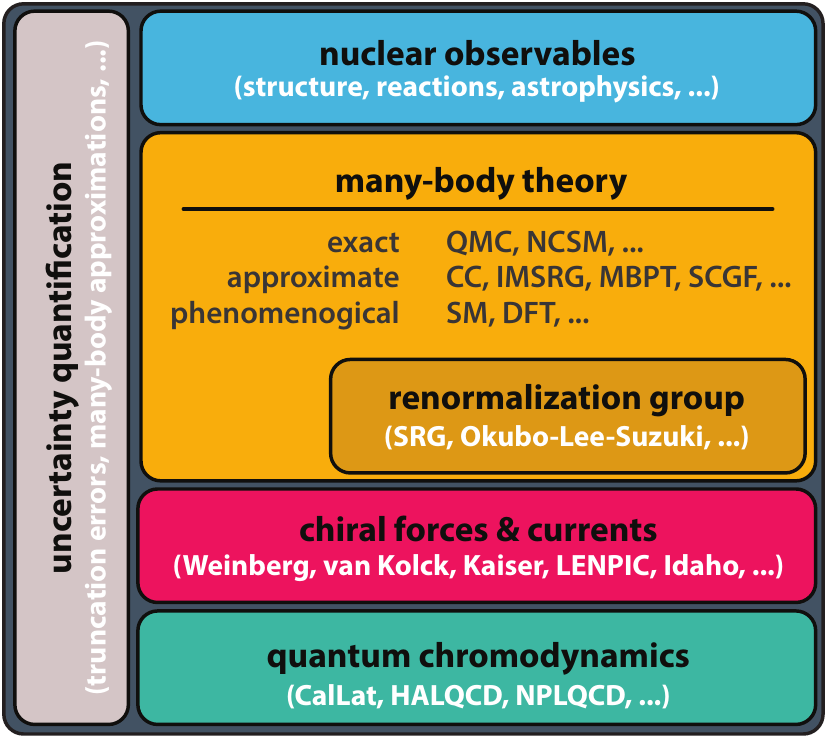}
	\end{center}
	\caption{
	Idealized workflow for \abinitio many-body calculations in modern nuclear theory (from the bottom to the top). See the main text for details.
	}
	\label{fig:mbt}
\end{figure}

Figure~\ref{fig:mbt} illustrates our conception of an idealized workflow for \abinitio many-body calculations in modern nuclear theory. The difficult task here is to make accurate predictions of nuclear observables with quantified and controllable theoretical uncertainties (blue box). 
Ideally, these predictions are directly derived from the theory of strong interactions, QCD, \eg, through lattice QCD calculations (turquoise box)~\cite{DRISCHLER2021103888}. 
But at the low-energy scales and finite densities relevant to nuclear physics, such direct calculations are, if possible at all, extremely challenging due to the nonperturbative nature of QCD and the fermion sign problem---even in the foreseeable future. 

Nuclear calculations with direct connection to QCD, none\-the\-less, have become within reach thanks to Weinberg's seminal papers in the early 1990s. 
Nowadays, chiral EFT with pion and nucleon degrees of freedom---sup\-ple\-mented with a truncation scheme known as Weinberg power counting---is the dominant microscopic approach to deriving nuclear forces and currents consistent with the symmetries of low-energy QCD (red box). 
The unresolved short-distance physics is expressed in terms of contact interactions with low-energy couplings fitted to experimental or lattice data.

\begin{figure}[tb]
	\begin{centering}
		\includegraphics[width=\linewidth]{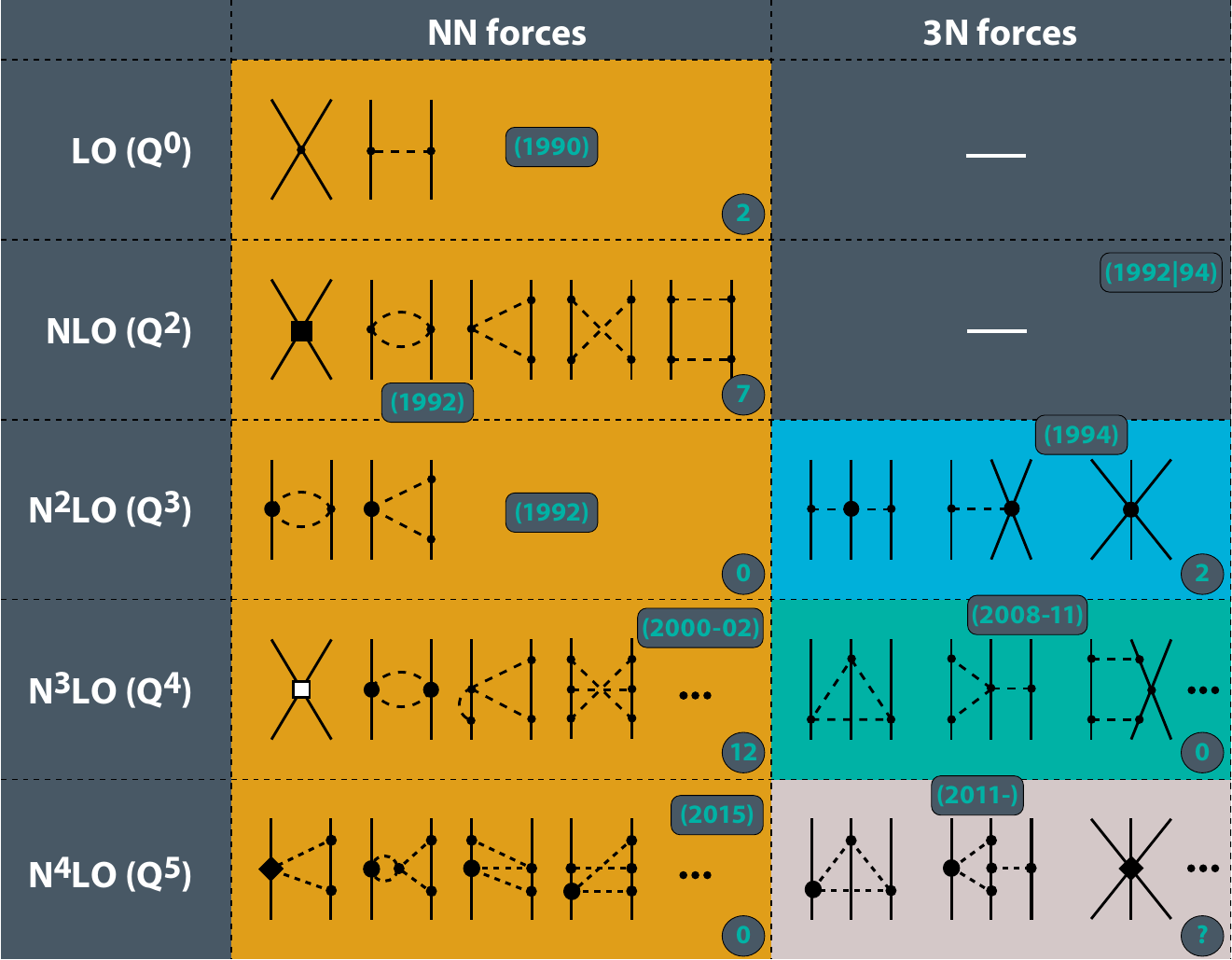}
	\end{centering}
	\caption{
	Hierarchy of chiral forces up to fifth order (or \NkLO{4}) in Weinberg power counting.
	Four-body forces at \NkLO{3} and \NkLO{4} are omitted for brevity.
	The chiral expansion is in powers of $Q=\max(p,m_\pi)/\Lambda_b$, with the typical momentum $p$ (or pion mass $m_\pi$) and EFT breakdown scale $\Lambda_b$.
	Nucleons are depicted by solid lines and pions by dashed lines.
	The boxed numbers represent the years in which the contributions were derived, and the circled numbers count the short-range contact low-energy couplings. See also Figure~4 in Ref.~\cite{Hebeler:2020ocj}.
	}
	\label{fig:eft_table}
\end{figure}

Figure~\ref{fig:eft_table} shows the hierarchy of chiral nuclear forces up to fifth order, or \NkLO{4}, in Weinberg power counting. 
Four-nucleon forces start contributing at \NkLO{3} but are omitted for brevity.
The boxed numbers represent the years in which the contributions were derived, and the circled numbers count the short-range contact low-energy couplings.
In recent years, significant progress has been made (\eg, by LENPIC~\cite{LENPICweb}) in constructing nuclear potentials up to a high order in the chiral expansion as well as exploring novel regularization schemes.

Given a chiral potential at some order and resolution scale, a computational framework is used to determine the nuclear observable of interest by solving the many-body Schr{\"o}dinger equation (light-orange box in Figure~\ref{fig:mbt}). 
Roughly, these many-body frameworks can be categorized into three categories (for a review, \eg, see Ref.~\cite{Hergert:2020bxy}): 
\begin{enumerate}
\item Methods that obtain (in principle) numerically exact solutions, such as Quantum Monte Carlo (QMC) methods and the No Core Shell Model (NCSM), and are therefore limited to light nuclei due to their exponential scaling. 

\item Approximate but systematically improvable methods based on many-body expansions, such as Coupled Cluster (CC) theory, In-Medium Similarity Re\-normalization Group (IMSRG), Many-Body Per\-tur\-bation Theory (MBPT), and Self-Con\-sis\-tent Green's Functions (SCGF) method. These methods scale polynomially and can reach well into the medi\-um-mass region. 

\item Phenomenological frameworks such as the Shell Mod\-el (SM) and Density Functional Theory (DFT), which together cover nearly the entire mass table and are the traditional workhorses for interpreting data and guiding experiments, but the connection to the underlying microscopic physics is unclear. 
\end{enumerate}

All of these frameworks have their own advantages, limitations, and numerical approximations associated. 
The strength of modern many-body theory lies in the ability to compare results for nuclear observables obtained in different frameworks to assess the underlying (many-body) approximations. 

The Renormalization Group (RG)~\cite{Bogner:2009bt} is a versatile tool to systematically modify the resolution scale of nuclear interactions while keeping observables invariant (dark-orange box in Figure~\ref{fig:mbt}).
Softening interactions using RG methods (as a preprocessing step) accelerates the rate of convergence of approximate many-body frameworks, but comes at the expense of inducing many-body forces---independent of the truncation level of the unevolved Hamiltonian.

In recent years, the similarity RG (SRG)~\cite{Bogner:2009bt} has been the method of choice for decoupling the high-to-low-momentum components of the nuclear interaction due to the strong short-range repulsion and tensor force.
Nuclear matrix elements are driven band- or block-diagonal in momentum space as the resolution scale is lowered through continuous infinitesimal unitary transformations dictated by the SRG flow equations.
In practice, induced many-body forces have to be truncated, \eg, at the three-body level. The SRG is only approximately unitary in these cases, and thus nuclear observables in systems with particle number greater than the truncation level will artificially depend on the resolution scale. One way to efficiently control the proliferation of induced many-body interactions is  to use normal-ordering with respect to an $A$-particle ``reference state''. This is the idea behind the IMSRG~\cite{Hergert:2015awm}---trun\-ca\-ting terms in the flow equations to $n$-body \emph{normal-ordered} operators defines the IMSRG($n$) approximation. In contrast to ``free-space'' SRG evolution, the IMSRG can be used as an \abinitio many-body method in and of itself. When starting from relatively soft NN and 3N interactions, the IMSRG($2$) already gives an excellent approximation for many nuclei, with controlled approximations to the IMSRG($3$) possible~\cite{Titusthesis,Heinz:2021xir}.

The Weinberg eigenvalue analysis has provided invaluable insights into  RG methods acting in different par\-tial-wave channels and nuclear potentials in general.
Weinberg (and others) developed this useful technique~\cite{PhysRev.131.440} in the early 1960s while he was at University of California, Berkeley, to understand how divergent Born series can be cured by introducing quasi-particles.
More recent applications of the Weinberg eigenvalue analysis include studies of the (rate of) convergence of perturbation expansions, both in free space (Born series)~\cite{Hoppe:2017lok} and in medium (MBPT)~\cite{Bogner:2009bt}, pairing instabilities~\cite{Ramanan:2007bb,Srinivas:2016kir}, and the construction of improved chiral NN potentials~\cite{Reinert:2017usi}.

Theoretical uncertainties arise at all layers shown in Figure~\ref{fig:mbt} and need to be quantified (gray box) for meaningful comparisons with, \eg, competing theories and constraints from nuclear experiment---where the latter typically are provided with a confidence interval.
The main sources of uncertainties in EFT calculations~\cite{Furnstahl:2014xsa} are due to the truncation of the EFT expansion at a finite order (\ie, the truncation error), 
fits of the low-energy couplings to experimental (or lattice) data, and 
approximations applied in the computational framework, \eg, to keep the numerical calculations tractable (see also Section~\ref{sec:uncertainties}).

%

\section{\Abinitio nuclear structure calculations} \label{sec:abinitio}

\begin{figure}[tb]
    \begin{center}
        \includegraphics[scale=1]{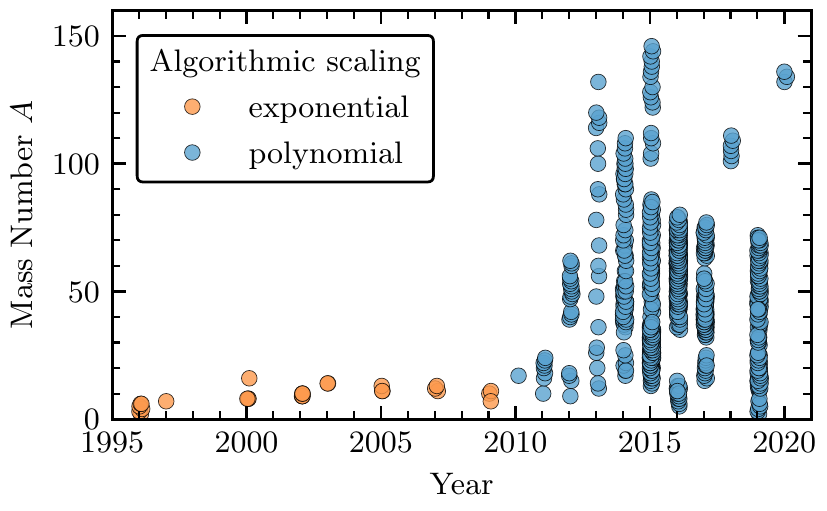}
    \end{center}
    \caption{
    Progress in microscopic nuclear structure calculations over the past 25 years (see also Ref.~\cite{Hagen:2015yea}).
    Data taken from Figure~1 in Ref.~\cite{Hergert:2020bxy}.
    See also the main text for details.
    }
    \label{fig:abintio_progress}
\end{figure}

Figure~\ref{fig:abintio_progress} summarizes the progress in microscopic nuclear structure calculations over the past 25 years. Until about 2010, the (approximately) linear increase in the highest mass number $A$ reachable in those calculations was determined by Moore's law and the exponential scaling of exact many-body methods (orange dots).
Since 2010, approximate but systematically improvable many-body methods 
with polynomial scaling in $A$ (see Section~\ref{sec:abinitio}) have pushed the frontier of state-of-the-art microscopic calculations to significantly higher mass numbers (blue dots)~\cite{Hergert:2020bxy}. 

Why did it take so long for these polynomially scaling methods to gain a foothold in nuclear many-body theory? The primary reason is that prior to the introduction of chiral EFT, phenomenological nuclear force models were rather hard, utilizing ultraviolet cutoffs or resolution scales on the order of several GeV or higher. Since these approximate many-body methods rely on expanding various quantities (\eg, two-body matrix elements of the NN potential) in a single-particle basis, it is essential that basis expansions converge rapidly for calculations to be tractable. For instance, the lowest non-trivial truncations of CC and IMSRG scale roughly as $N^6$, where $N$ is the number of included  single-particle orbitals. Demanding that the single-par\-ti\-cle basis is sufficiently extended in coordinate space to capture the spatial extent of the nucleus, and sufficiently extended in momentum space to capture relevant momentum modes up to the resolution scale in the nuclear potentials, one can use semi-classical arguments to show that $N$ scales as $\Lambda^3$, with the resolution scale~$\Lambda$. 

The bottom line is that even a modest reduction in $\Lambda$, such as that in going from the hard phenomenological nuclear force models to the softer chiral interactions, can have a profound impact on the viability of these approximate many-body methods. The impressive progress shown in Figure~\ref{fig:abintio_progress} would not have been possible without the computational simplifications afforded by soft (and even softer RG-evolved) chiral interactions. That said, it is ironic that the softness that has been so central to the many-body progress is a consequence of inconsistencies in Weinberg power counting that prevent one from taking the cutoff to larger values---see the discussion in Ref.~\cite{vanKolck:2021rqu}.

\begin{figure}[tb]
    \begin{center}
        \includegraphics[scale=1]{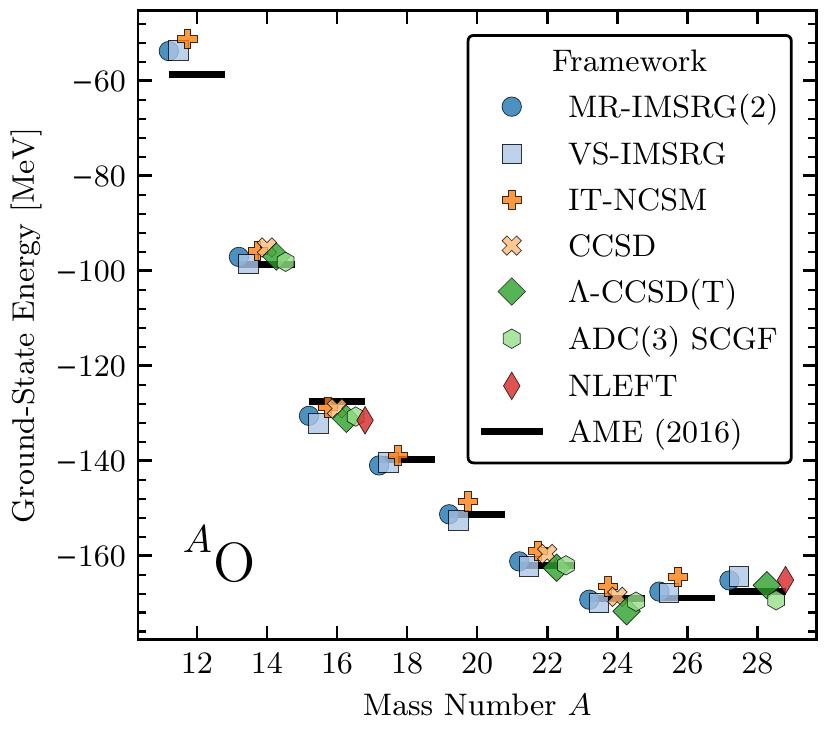}
    \end{center}
    \caption{
    Predictions for the ground-state energies of the oxygen isotopes obtained using several many-body frameworks (symbols). All calculations are based on the same low-momentum NN and 3N interaction, apart from those obtained in nuclear lattice EFT (NLEFT).
    For details see the discussion of Figure~5 in Ref.~\cite{Hergert:2020bxy}, from which the data is taken.
    }
    \label{fig:oxygen}
\end{figure}

The advancement of \abinitio theory well into the medium-mass region is an impressive feat, but the physics value lies in the fact that such calculations are becoming increasingly precise. Figure~\ref{fig:oxygen} is one such illustration, where a wide variety of many-body methods starting from the same SRG-evolved chiral NN and 3N potential 
are in good agreement with experiment and each other for the oxygen isotopes. Note that ``good agreement'' in the present context is somewhat ill-defined since the calculations do not come with error bars reflecting the uncertainties in the input chiral interactions and the subsequent many-body approximations.   

For the time being, we content ourselves with the following comments. 
First, while the many-body truncation errors of the different methods in Figure~\ref{fig:oxygen} are not rigorously quantified, there is numerical evidence that they are significantly smaller than the uncertainties associated with the input chiral interactions. Therefore, the urgent task from the perspective of UQ is to propagate the chiral EFT uncertainties through the many-body calculations. See Ref.~\cite{Maris:2020qne} where impressive progress has been made in carrying out a comprehensive error analysis for \abinitio calculations of light nuclei.  
Second, the ``good agreement'' with experiment tends to degrade substantially for total energies and radii as one moves to heavier nuclei for nearly all chiral NN and 3N interactions on the market. This deficiency is the primary obstacle to precision calculations in microscopic nuclear structure theory, though as discussed in Section~\ref{sec:uncertainties}, the recent development of accurate and efficient emulators for UQ holds much promise in addressing this issue. 

The phenomenological shell model has enjoyed tre\-mendous success as the main workhorse of nuclear structure theory, but it requires an abundance of data in the mass region of interest in order to pin down the effective valence Hamiltonian. As nuclear physics shifts its focus to the study of rare isotopes in unexplored regions of the nuclear chart, this places a strain on data-driven approaches like the phenomenological shell model, requiring microscopic approaches to play a greater role. 

Along these lines, an important recent development in \abinitio theory is the controlled derivation of shell model Hamiltonians and effective operators, starting from chiral NN and 3N interactions~\cite{Bogner:2014baa, Stroberg:2016ung,Sun:2021ffi}. While there had been previous efforts that enjoyed some quantitative success dating back to the 1960s, they were plagued with convergence issues that trace back to the high resolution scales of the pre-chiral EFT NN interactions. The shell model is an intrinsically low-resolution description since the underlying picture of a dominant attractive mean field plus a weak residual interaction implies a decoupling of high- and low-momentum modes in the Hamiltonian. In this sense, chiral NN and 3N interactions give a much closer starting point to this picture, and the subsequent transformations to decouple the valence nucleons are easier to treat with controlled approximations.  

Another difficulty with the older efforts to derive shell model Hamiltonians is that very little was known about the impact of 3N interactions. This was in part because of the computational difficulties of including them, but also because there was no underlying framework that said they needed to be included at a given level of description as in chiral EFT---they were merely asserted to be small and usually neglected. Indeed, the empirical adjustments that need to be made to the older microscopically derived shell model Hamiltonians, such as the ``monopole correction'' of Zuker, are thought to mock up the effects of neglected 3N forces. As discussed in Ref.~\cite{Stroberg:2019mxo}, the valence-space IMSRG (VS--IMSRG) explicitly demonstrates that this is indeed the case. More generally, the inclusion of chiral 3N interactions has been shown to play a crucial role in accurately describing the evolution of structure along isotopic chains~\cite{Hebeler:2015hla}.    

Perhaps one of the more impressive applications of the new generation of \abinitio shell model Hamiltonians is the recent work of Stroberg~\etal~\cite{Stroberg:2019gmc} that computed ground-state and separation energies of nearly 700 isotopes, in what is arguably the first \abinitio mass table calculation up to the iron isotopes. Performing a Bayesian linear regression analysis for the residual of the various separation energies, Stroberg~\etal were then able to predict the location of the neutron and proton driplines probabilistically.
They used the softest of the interactions by Hebeler~\etal~\cite{Hebeler:2010xb}, where the chiral NN interaction is softened by SRG evolution, and the leading chiral 3N forces are fit to reproduce the \isotope[3]{H} ground-state energy and \isotope[4]{He} charge radius. While inconsistent from an EFT perspective, this interaction gives a strikingly good description of nuclear matter saturation properties~\cite{Hebeler:2010xb,Drischler:2017wtt} and ground- and excited-state energies well into the medium-mass region~\cite{Simonis:2017dny}, and is informally referred to as the ``magic'' interaction. Radii likewise are closer to experiment than most chiral interactions, though the systematic underprediction remains.

We have only scratched the surface of a very deep and rapidly evolving area of research. Recent advances not covered here include the explanation of the quenching of $g_A$ in medium-mass nuclei as an interplay of correlation effects and two-body currents~\cite{Gysbers:2019uyb}, the first \abinitio calculations of neutrinoless double beta decay matrix elements~\cite{Yao:2019rck,Belley:2020ejd}, and many more.



\section{Quantification of theoretical uncertainties} \label{sec:uncertainties}

Quantifying theoretical uncertainties is an integral component of the \abinitio workflow depicted in Figure~\ref{fig:mbt}. 
The focus has recently shifted from probing uncertainties by simple (somewhat arbitrary) parameter variations toward systematic studies of EFT truncation errors~\cite{Furnstahl:2014xsa}.
Epelbaum~\etal~\cite{Epelbaum:2014efa,Epelbaum:2014sza} introduced an implementation of the standard EFT uncertainty that assumes that the truncation error of an observable is uncorrelated and dominated by the first chiral order in Figure~\ref{fig:eft_table} not included in the many-body calculations. 
This ``EKM uncertainty'' has been extensively stud\-ied in \abinitio calculations of atomic nuclei~\cite{Binder:2015mbz,Epelbaum:2019kcf} and nuclear matter~\cite{Drischler:2017wtt,Lonardoni:2019ypg}.
Furthermore, the Bayesian UQ: Errors in Your EFT (BUQEYE) Collaboration~\cite{BUQEYEweb} has been developing advanced statistical methods for rigorous quantification and propagation of (correlated) theoretical uncertainties, including the inference of to-all-orders EFT truncation errors and breakdown scales, via order-by-order calculations, \eg, of scattering and nuclear matter observables~\cite{Melendez:2017phj,Melendez:2019izc,Drischler:2020yad}.

The progress in applying statistical methods, such as Bayesian parameter estimation~\cite{Wesolowski:2018lzj}, model comparison~\cite{Phillips:2020dmw}, and sensitivity analysis~\cite{Ekstrom:2019lss}, is apparent in the successful Information and Statistics in Nuclear Experiment and Theory (ISNET) workshops~\cite{0954-3899-42-3-030301,ISNETweb}. 
Since 2012 these annual workshops bring nuclear physicists and statisticians together to learn and discuss statistical methods and their applications to nuclear physics problems in an inclusive scientific environment. 
The pedagogical lectures during the workshops, hos\-ted by the Bayesian Analysis for Nuclear Dynamics (BAND) Framework Collaboration~\cite{BANDframework},  are an excellent way of getting starting with Bayesian methods.

\begin{figure}[tb]
    \begin{center}
        \includegraphics[scale=1]{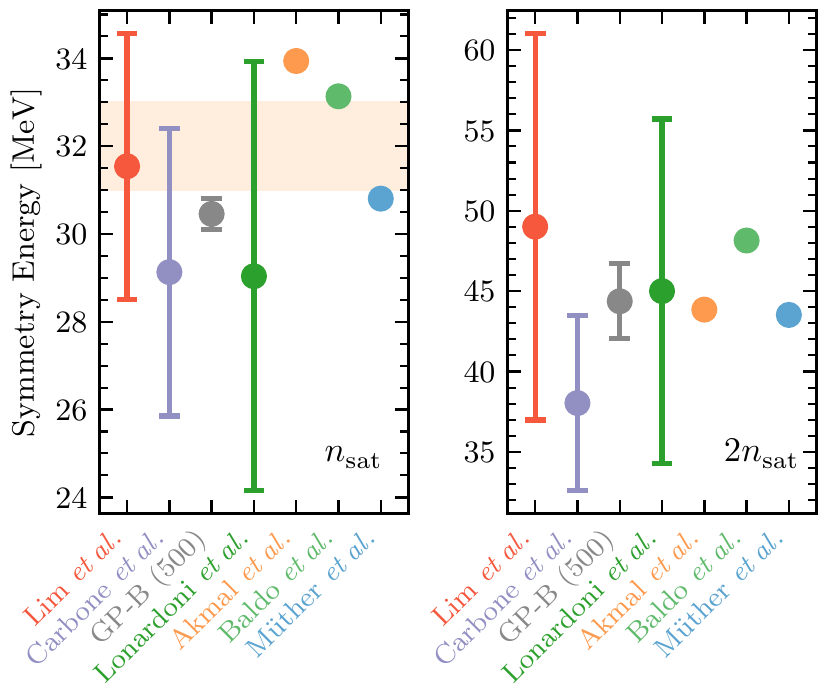}
    \end{center}
    \caption{
    Constraints on the nuclear symmetry energy at $n = \nsat$ (left panel) and $2\nsat$ (right panel) from microscopic nuclear matter calculations with chiral NN and 3N interactions (error bars) in comparison with older phenomenological calculations (dots).
    The canonical value of $\nsat = 0.16 \fmiq$ is used.
    Data were taken from Figure~4 in Ref.~\cite{Drischler:2021kxf} and are based on the recent chiral EFT calculations by 
    Lim~\& Holt~\cite[MBPT]{lim18}, 
    Carbone~\etal~\cite[SCGF]{Carbone:2014mja}, 
    Lonardoni~\etal~\cite[QMC]{Lonardoni:2019ypg} ($E_{\openone}$ parametrization), and
    Drischler~\etal~\cite[MBPT, ``GP--B(500)'']{Drischler:2020hwi,Drischler:2017wtt}, as well as the phenomenological calculations by
    Akmal~\etal~\cite{akmal98}, 
    Baldo~\etal~\cite{baldo97}, and
    M{\"u}ther~\etal~\cite{muether87}. 
    The orange band depicts the model average of energy density functionals by Reinhard~\etal~\cite{Reinhard:2021utv} and serves only as a reference.
    }
    \label{fig:Esym}
\end{figure}

Let us use nuclear matter as an example to highlight UQ in chiral EFT calculations.
A recent review of the nuclear EOS with UQ and astrophysical applications can be found in Ref.~\cite{Drischler:2021kxf}.
Figure~\ref{fig:Esym} compares microscopic constraints (error bars) on the nuclear symmetry energy---a key quantity in nuclear (astro)physics---obtained with different many-body frameworks and chiral NN and 3N interactions at $n=\nsat$ (left panel) and $2\nsat$ (right panel).
See the caption of Figure~\ref{fig:Esym} for more details on the calculations.
Only because uncertainties were estimated can one judge how well (in this case) the symmetry energy is determined microscopically: overall, within the relatively large uncertainties, the different constraints are consistent with one another, and the symmetry energy is predicted in the range $E_\text{sym}(\nsat) \approx 24.1-34.6 \MeV$ and $E_\text{sym}(2\nsat) \approx 32.5 - 61.1 \MeV$, respectively, as shown in Figure~\ref{fig:Esym}.
On the other hand, the older phenomenological calculations---how accurate those may or may not be---only provide point estimates (dots), while the underlying uncertainties are unknown, and thus do not allow for such an interpretation.
However, we stress that the methods used for estimating the uncertainties in Figure~\ref{fig:Esym} are conceptually different, ranging from parameter variations in the nuclear interactions to truncation error studies.
Rigorously quantifying all theoretical uncertainties available within a comprehensive statistical framework to constrain the nuclear EOS (not only the symmetry energy) from different many-body methods and chiral interactions is an important avenue for the future. This will require the development of improved order-by-order NN and 3N interactions~\cite{Huther:2019ont}.
The Bayesian framework for quantifying EFT truncation errors introduced by the BUQEYE Collaboration~\cite{Drischler:2020hwi,Drischler:2020yad} (and applied to the structure of neutron stars~\cite{Drischler:2020fvz}) is the first step in this direction.

These full Bayesian analyses typically are computationally demanding and thus prohibitively slow in most cases---or so one might have thought. 
However, emulators for nuclear observables---where applicable---have overcome this computational limitation by providing highly accurate approximations to exact solutions (\eg, of the Schr{\"o}dinger equation) fast and reliably. 
But applications of these powerful new tools reach way beyond that.
Several methods have been used to implement emulators, with eigenvector continuation (EC)~\cite{Frame:2017fah} being one of the promising methods. 
For instance, EC-driven emulators have been demon\-strat\-ed to be remarkably efficient in emulating, \eg, ground-state energies and charge radii~\cite{Ekstrom:2019lss,Konig:2019adq,Wesolowski:2021cni}.
Furthermore, the EC concept has been extended to emulate two-body scattering observables using the variational principles due to Kohn~\cite{Furnstahl:2020abp,Drischler:2021qoy} and Newton~\cite{Melendez:2021lyq}, respectively, while the first application to three-body scattering---where the emulators' efficacies are put to a real display---is already encouraging for systematic studies of emulated three- and higher-body scattering observables in the future~\cite{Zhang:2021xx}.
A fast \& accurate emulator for few-body bound-state properties was recently key to constructing the first set of order-by-order chiral NN and 3N interactions (up to \NkLO{2}) with theoretical uncertainties fully quantified~\cite{Wesolowski:2021cni}. 
The task now is to bridge the developments in Bayesian methods and emulator technology toward a full UQ in microscopic calculations of nuclear matter and atomic nuclei across the nuclear chart.

For nuclear matter, such a rigorous UQ has never been more important than today given the great variety of empirical constraints on the nuclear EOS recently obtained or anticipated soon.
These constraints include 
tidal deformabilities of neutron stars inferred from direct grav\-i\-ta\-tional-wave detection by the LIGO-Virgo scientific collaboration (now joined by the KAGRA observatory)~\cite{Abbott:2018exr,De:2018uhw,Capano:2019eae}, 
simultaneous mass-radius measurements of neutron stars by the NICER telescope~\cite{Miller:2021qha,Riley:2021pdl}, and 
neutron skins, \eg, measured by the second \isotope[208]{Pb} Radius EXperiment (PREX--II) at Jefferson Lab~\cite{PREX:2021umo,Reed:2021nqk}.
Statistically meaningful comparisons of EOS constraints from observation and experiment with chiral EFT predictions, especially at $n \approx 1-2\,\nsat$, shed light on strongly interacting matter under extreme conditions and, through the \abinitio workflow in Figure~\ref{fig:mbt}, also on the underlying microscopic nuclear interactions.
This has been, and will continue to be, an exciting era for nuclear (astro)physics---about 30 years after Weinberg's seminal papers were published.

\section{Conclusion} \label{sec:summary}

Celebrating Weinberg's scientific achievements, we argued that nuclear physics has recently entered a precision era thanks to chiral EFT, for which his seminal papers~\cite{Weinberg:1990rz,Weinberg:1991um,Weinberg:1992yk} in the early 1990s laid the groundwork.
Specifically, we pointed out chiral EFT's advantages over previous phenomenological nuclear approaches and shared our perspective on some of the recent advances made in \abinitio calculations of nuclear structure and nuclear matter observables, as well as the quantification of EFT truncation errors. 

With increasingly accurate and efficient many-body frameworks available, however, long-lasting formal issues in chiral EFT (\eg, with the power counting) resurface as the uncertainties from the chiral interactions gradually become more important than those from the many-body frameworks. 
These issues were discussed in the recent INT Program Nuclear Structure at the Crossroads (INT--19--2a) and summarized in the organizers' contribution to this special issue~\cite{Furnstahl:2021rfk}. 
From this perspective, despite all the progress in many-body calculations because of chiral EFT, it seems as if nuclear theory is indeed at the crossroads with unknown roads ahead. Bayesian methods combined with emulators for nuclear observables (see Section~\ref{sec:uncertainties}) will provide invaluable guidance along the way, though.

In his contribution to this special issue~\cite{vanKolck:2021rqu}, Bira van Kolck vividly tells an anecdote of encountering an agitated Steven Weinberg after he had called Gerry Brown sometime in 1990. 
Weinberg wanted to pick Gerry's brain because he knew him to be one of the leading experts on nuclear forces and nuclear physics at that time. 
The phone call must have been confusing and agitating for both sides. Despite the fact that they shared a common conviction of the central role of chiral symmetry in understanding nuclear forces---indeed, Gerry and collaborators wrote several papers in which they forcefully made this point ten years prior---they spoke different scientific languages and had different points of emphasis that clashed with each other. Weinberg was frustrated because Gerry stubbornly kept on emphasizing fitting data and the importance of keeping the rho meson as an explicit degree of freedom, in part because it is needed to cancel the ``unphysical'' large attractive tensor force of the one-pion exchange potential at short distances. And Gerry was equally frustrated that Weinberg refused to treat the rho meson as a low-energy degree of freedom due to its large mass near the EFT breakdown scale.

About ten years later, when one of us (S.B.) was a wide-eyed graduate student at Stony Brook together with Achim Schwenk, who would turn out to be the thesis advisor for another one of us (C.D.), Gerry regaled us about the infamous phone call. It's worth mentioning that Gerry told the story with a wry smile on his face, and seemed to take special delight in the fact that he had made Weinberg so mad!\footnote{Gerry was such a masterful storyteller that S.B. always wondered if he was taking a bit of artistic license for comic effect in his recollection of the phone call. Several years later---it might have been over pints in Trento or Seattle during some long-forgotten conference---Bira ended any lingering doubts by basically telling the anecdote that is in his contribution.} It's also worth mentioning that Gerry was, apart from his unwavering feelings about the rho meson, a strong supporter of the EFT ideology in general, and he had tremendous respect for Weinberg's seminal work in this area. We can only speculate what Gerry had exactly in mind during this contentious phone call. But as Bira wrote in his contribution~\cite{vanKolck:2021rqu}, it may very well be that Gerry's deep intuition for nuclear forces let him anticipate some of the issues that we are still facing in chiral EFT with Weinberg power counting.
However, beyond speculation we are confident that, no matter what quantum state Steven Weinberg and Gerry Brown may be in together now, they will not run out of interesting physics questions to discuss anytime soon. In memory of Steven Weinberg (1933--2021) and Gerry Brown (1926--2013).


\begin{acknowledgements}

We thank our colleagues and collaborators who have helped shape our understanding of nuclear EFT and many-body theory.
We are also grateful to Alejandro Kievsky for the kind invitation to contribute to this special issue and Heiko Hergert for providing us with the data for Figures~\ref{fig:abintio_progress} and~\ref{fig:oxygen}.
This material is based upon work supported by the U.S. Department of Energy, Office of Science, Office of Nuclear Physics, under the FRIB Theory Alliance award DE-SC0013617, and the National Science Foundation awards PHY-2013047 and PHY-1713901.

\end{acknowledgements}

\bibliographystyle{apsrev4-1}
\bibliography{refs,bayesian_refs}   

\begin{thebibliography}{88}%
\makeatletter
\providecommand \@ifxundefined [1]{%
 \@ifx{#1\undefined}
}%
\providecommand \@ifnum [1]{%
 \ifnum #1\expandafter \@firstoftwo
 \else \expandafter \@secondoftwo
 \fi
}%
\providecommand \@ifx [1]{%
 \ifx #1\expandafter \@firstoftwo
 \else \expandafter \@secondoftwo
 \fi
}%
\providecommand \natexlab [1]{#1}%
\providecommand \enquote  [1]{``#1''}%
\providecommand \bibnamefont  [1]{#1}%
\providecommand \bibfnamefont [1]{#1}%
\providecommand \citenamefont [1]{#1}%
\providecommand \href@noop [0]{\@secondoftwo}%
\providecommand \href [0]{\begingroup \@sanitize@url \@href}%
\providecommand \@href[1]{\@@startlink{#1}\@@href}%
\providecommand \@@href[1]{\endgroup#1\@@endlink}%
\providecommand \@sanitize@url [0]{\catcode `\\12\catcode `\$12\catcode
  `\&12\catcode `\#12\catcode `\^12\catcode `\_12\catcode `\%12\relax}%
\providecommand \@@startlink[1]{}%
\providecommand \@@endlink[0]{}%
\providecommand \url  [0]{\begingroup\@sanitize@url \@url }%
\providecommand \@url [1]{\endgroup\@href {#1}{\urlprefix }}%
\providecommand \urlprefix  [0]{URL }%
\providecommand \Eprint [0]{\href }%
\providecommand \doibase [0]{http://dx.doi.org/}%
\providecommand \selectlanguage [0]{\@gobble}%
\providecommand \bibinfo  [0]{\@secondoftwo}%
\providecommand \bibfield  [0]{\@secondoftwo}%
\providecommand \translation [1]{[#1]}%
\providecommand \BibitemOpen [0]{}%
\providecommand \bibitemStop [0]{}%
\providecommand \bibitemNoStop [0]{.\EOS\space}%
\providecommand \EOS [0]{\spacefactor3000\relax}%
\providecommand \BibitemShut  [1]{\csname bibitem#1\endcsname}%
\let\auto@bib@innerbib\@empty
\bibitem [{\citenamefont {{{The University of Texas at Austin: UT
  News}}}(2021)}]{WeinbergUTAus}%
  \BibitemOpen
  \bibfield  {author} {\bibinfo {author} {\bibnamefont {{{The University of
  Texas at Austin: UT News}}}},\ }\href@noop {} {} (\bibinfo {year} {2021}),\
  \bibinfo {note}
  {\href{https://news.utexas.edu/2021/07/24/ut-austin-mourns-death-of-world-renowned-physicist-steven-weinberg/}{https://news.utexas.edu/2021/07/24/ut-austin-mourns-death-of-world-renowned-physicist-steven-weinberg/}}\BibitemShut
  {NoStop}%
\bibitem [{\citenamefont {Weinberg}(1990)}]{Weinberg:1990rz}%
  \BibitemOpen
  \bibfield  {author} {\bibinfo {author} {\bibfnamefont {S.}~\bibnamefont
  {Weinberg}},\ }\href@noop {} {\bibfield  {journal} {\bibinfo  {journal}
  {Phys. Lett. B}\ }\textbf {\bibinfo {volume} {251}},\ \bibinfo {pages} {288}
  (\bibinfo {year} {1990})}\BibitemShut {NoStop}%
\bibitem [{\citenamefont {Weinberg}(1991)}]{Weinberg:1991um}%
  \BibitemOpen
  \bibfield  {author} {\bibinfo {author} {\bibfnamefont {S.}~\bibnamefont
  {Weinberg}},\ }\href@noop {} {\bibfield  {journal} {\bibinfo  {journal}
  {Nucl. Phys. B}\ }\textbf {\bibinfo {volume} {363}},\ \bibinfo {pages} {3}
  (\bibinfo {year} {1991})}\BibitemShut {NoStop}%
\bibitem [{\citenamefont {Weinberg}(1992)}]{Weinberg:1992yk}%
  \BibitemOpen
  \bibfield  {author} {\bibinfo {author} {\bibfnamefont {S.}~\bibnamefont
  {Weinberg}},\ }\href@noop {} {\bibfield  {journal} {\bibinfo  {journal}
  {Phys. Lett. B}\ }\textbf {\bibinfo {volume} {295}},\ \bibinfo {pages} {114}
  (\bibinfo {year} {1992})},\ \Eprint {http://arxiv.org/abs/hep-ph/9209257}
  {hep-ph/9209257} \BibitemShut {NoStop}%
\bibitem [{\citenamefont {{\textsc{Scimeter}: keyword cloud generator based on
  data from the arXiv preprint server}}(2020)}]{Scimeterweb}%
  \BibitemOpen
  \bibfield  {author} {\bibinfo {author} {\bibnamefont {{\textsc{Scimeter}:
  keyword cloud generator based on data from the arXiv preprint server}}},\
  }\href {https://scimeter.org/} {} (\bibinfo {year} {2020}),\ \bibinfo {note}
  {\url{https://scimeter.org/} (maintained by Frankfurt Institute for Advanced
  Studies (FIAS))}\BibitemShut {NoStop}%
\bibitem [{\citenamefont {Epelbaum}(2016)}]{Epelbaum:2015pfa}%
  \BibitemOpen
  \bibfield  {author} {\bibinfo {author} {\bibfnamefont {E.}~\bibnamefont
  {Epelbaum}},\ }\href {\doibase 10.22323/1.253.0014} {\bibfield  {journal}
  {\bibinfo  {journal} {PoS CD}\ }\textbf {\bibinfo {volume} {15}},\ \bibinfo
  {pages} {014} (\bibinfo {year} {2016})},\ \Eprint
  {http://arxiv.org/abs/1510.07036} {arXiv:1510.07036} \BibitemShut {NoStop}%
\bibitem [{\citenamefont {Epelbaum}\ \emph {et~al.}(2020)\citenamefont
  {Epelbaum}, \citenamefont {Krebs},\ and\ \citenamefont
  {Reinert}}]{Epelbaum:2019kcf}%
  \BibitemOpen
  \bibfield  {author} {\bibinfo {author} {\bibfnamefont {E.}~\bibnamefont
  {Epelbaum}}, \bibinfo {author} {\bibfnamefont {H.}~\bibnamefont {Krebs}}, \
  and\ \bibinfo {author} {\bibfnamefont {P.}~\bibnamefont {Reinert}},\ }\href
  {\doibase 10.3389/fphy.2020.00098} {\bibfield  {journal} {\bibinfo  {journal}
  {Front. Phys.}\ }\textbf {\bibinfo {volume} {8}},\ \bibinfo {pages} {98}
  (\bibinfo {year} {2020})},\ \Eprint {http://arxiv.org/abs/1911.11875}
  {arXiv:1911.11875} \BibitemShut {NoStop}%
\bibitem [{\citenamefont {Hammer}\ \emph {et~al.}(2020)\citenamefont {Hammer},
  \citenamefont {K\"onig},\ and\ \citenamefont {van Kolck}}]{Hammer:2019poc}%
  \BibitemOpen
  \bibfield  {author} {\bibinfo {author} {\bibfnamefont {H.-W.}\ \bibnamefont
  {Hammer}}, \bibinfo {author} {\bibfnamefont {S.}~\bibnamefont {K\"onig}}, \
  and\ \bibinfo {author} {\bibfnamefont {U.}~\bibnamefont {van Kolck}},\ }\href
  {\doibase 10.1103/RevModPhys.92.025004} {\bibfield  {journal} {\bibinfo
  {journal} {Rev. Mod. Phys.}\ }\textbf {\bibinfo {volume} {92}},\ \bibinfo
  {pages} {025004} (\bibinfo {year} {2020})},\ \Eprint
  {http://arxiv.org/abs/1906.12122} {arXiv:1906.12122} \BibitemShut {NoStop}%
\bibitem [{\citenamefont {Machleidt}\ and\ \citenamefont
  {Entem}(2011)}]{Machleidt:2011zz}%
  \BibitemOpen
  \bibfield  {author} {\bibinfo {author} {\bibfnamefont {R.}~\bibnamefont
  {Machleidt}}\ and\ \bibinfo {author} {\bibfnamefont {D.~R.}\ \bibnamefont
  {Entem}},\ }\href {\doibase 10.1016/j.physrep.2011.02.001} {\bibfield
  {journal} {\bibinfo  {journal} {Phys. Rept.}\ }\textbf {\bibinfo {volume}
  {503}},\ \bibinfo {pages} {1} (\bibinfo {year} {2011})},\ \Eprint
  {http://arxiv.org/abs/1105.2919} {arXiv:1105.2919} \BibitemShut {NoStop}%
\bibitem [{\citenamefont {Epelbaum}\ \emph {et~al.}(2009)\citenamefont
  {Epelbaum}, \citenamefont {Hammer},\ and\ \citenamefont
  {Mei{\ss}ner}}]{Epelbaum:2008ga}%
  \BibitemOpen
  \bibfield  {author} {\bibinfo {author} {\bibfnamefont {E.}~\bibnamefont
  {Epelbaum}}, \bibinfo {author} {\bibfnamefont {H.-W.}\ \bibnamefont
  {Hammer}}, \ and\ \bibinfo {author} {\bibfnamefont {U.-G.}\ \bibnamefont
  {Mei{\ss}ner}},\ }\href {\doibase 10.1103/RevModPhys.81.1773} {\bibfield
  {journal} {\bibinfo  {journal} {Rev. Mod. Phys.}\ }\textbf {\bibinfo {volume}
  {81}},\ \bibinfo {pages} {1773} (\bibinfo {year} {2009})},\ \Eprint
  {http://arxiv.org/abs/0811.1338} {arXiv:0811.1338} \BibitemShut {NoStop}%
\bibitem [{\citenamefont {Hebeler}(2021)}]{Hebeler:2020ocj}%
  \BibitemOpen
  \bibfield  {author} {\bibinfo {author} {\bibfnamefont {K.}~\bibnamefont
  {Hebeler}},\ }\href {\doibase 10.1016/j.physrep.2020.08.009} {\bibfield
  {journal} {\bibinfo  {journal} {Phys. Rept.}\ }\textbf {\bibinfo {volume}
  {890}},\ \bibinfo {pages} {1} (\bibinfo {year} {2021})},\ \Eprint
  {http://arxiv.org/abs/2002.09548} {arXiv:2002.09548 [nucl-th]} \BibitemShut
  {NoStop}%
\bibitem [{\citenamefont {Hebeler}\ \emph {et~al.}(2015)\citenamefont
  {Hebeler}, \citenamefont {Holt}, \citenamefont {Menendez},\ and\
  \citenamefont {Schwenk}}]{Hebeler:2015hla}%
  \BibitemOpen
  \bibfield  {author} {\bibinfo {author} {\bibfnamefont {K.}~\bibnamefont
  {Hebeler}}, \bibinfo {author} {\bibfnamefont {J.~D.}\ \bibnamefont {Holt}},
  \bibinfo {author} {\bibfnamefont {J.}~\bibnamefont {Menendez}}, \ and\
  \bibinfo {author} {\bibfnamefont {A.}~\bibnamefont {Schwenk}},\ }\href
  {\doibase 10.1146/annurev-nucl-102313-025446} {\bibfield  {journal} {\bibinfo
   {journal} {Ann. Rev. Nucl. Part. Sci.}\ }\textbf {\bibinfo {volume} {65}},\
  \bibinfo {pages} {457} (\bibinfo {year} {2015})},\ \Eprint
  {http://arxiv.org/abs/1508.06893} {arXiv:1508.06893} \BibitemShut {NoStop}%
\bibitem [{\citenamefont {Bogner}\ \emph {et~al.}(2010)\citenamefont {Bogner},
  \citenamefont {Furnstahl},\ and\ \citenamefont {Schwenk}}]{Bogner:2009bt}%
  \BibitemOpen
  \bibfield  {author} {\bibinfo {author} {\bibfnamefont {S.~K.}\ \bibnamefont
  {Bogner}}, \bibinfo {author} {\bibfnamefont {R.~J.}\ \bibnamefont
  {Furnstahl}}, \ and\ \bibinfo {author} {\bibfnamefont {A.}~\bibnamefont
  {Schwenk}},\ }\href {\doibase 10.1016/j.ppnp.2010.03.001} {\bibfield
  {journal} {\bibinfo  {journal} {Prog. Part. Nucl. Phys.}\ }\textbf {\bibinfo
  {volume} {65}},\ \bibinfo {pages} {94} (\bibinfo {year} {2010})},\ \Eprint
  {http://arxiv.org/abs/0912.3688} {arXiv:0912.3688} \BibitemShut {NoStop}%
\bibitem [{\citenamefont {Furnstahl}\ \emph {et~al.}(2014)\citenamefont
  {Furnstahl}, \citenamefont {Papenbrock},\ and\ \citenamefont
  {More}}]{Furnstahl:2013vda}%
  \BibitemOpen
  \bibfield  {author} {\bibinfo {author} {\bibfnamefont {R.~J.}\ \bibnamefont
  {Furnstahl}}, \bibinfo {author} {\bibfnamefont {T.}~\bibnamefont
  {Papenbrock}}, \ and\ \bibinfo {author} {\bibfnamefont {S.~N.}\ \bibnamefont
  {More}},\ }\href {\doibase 10.1103/PhysRevC.89.044301} {\bibfield  {journal}
  {\bibinfo  {journal} {Phys. Rev. C}\ }\textbf {\bibinfo {volume} {89}},\
  \bibinfo {pages} {044301} (\bibinfo {year} {2014})},\ \Eprint
  {http://arxiv.org/abs/1312.6876} {arXiv:1312.6876 [nucl-th]} \BibitemShut
  {NoStop}%
\bibitem [{\citenamefont {Lattimer}(2021)}]{Lattimer2021annrev}%
  \BibitemOpen
  \bibfield  {author} {\bibinfo {author} {\bibfnamefont {J.}~\bibnamefont
  {Lattimer}},\ }\href {\doibase 10.1146/annurev-nucl-102419-124827} {\bibfield
   {journal} {\bibinfo  {journal} {Annu. Rev. Nucl. Part. Sci.}\ }\textbf
  {\bibinfo {volume} {71}},\ \bibinfo {pages} {433–64} (\bibinfo {year}
  {2021})}\BibitemShut {NoStop}%
\bibitem [{\citenamefont {Drischler}\ \emph
  {et~al.}(2021{\natexlab{a}})\citenamefont {Drischler}, \citenamefont {Holt},\
  and\ \citenamefont {Wellenhofer}}]{Drischler:2021kxf}%
  \BibitemOpen
  \bibfield  {author} {\bibinfo {author} {\bibfnamefont {C.}~\bibnamefont
  {Drischler}}, \bibinfo {author} {\bibfnamefont {J.~W.}\ \bibnamefont {Holt}},
  \ and\ \bibinfo {author} {\bibfnamefont {C.}~\bibnamefont {Wellenhofer}},\
  }\href {\doibase 10.1146/annurev-nucl-102419-041903} {\bibfield  {journal}
  {\bibinfo  {journal} {Annu. Rev. Nucl. Part. Sci.}\ }\textbf {\bibinfo
  {volume} {71}},\ \bibinfo {pages} {403} (\bibinfo {year}
  {2021}{\natexlab{a}})},\ \Eprint {http://arxiv.org/abs/2101.01709}
  {arXiv:2101.01709} \BibitemShut {NoStop}%
\bibitem [{\citenamefont {Tews}(2020)}]{Tews:2020wrl}%
  \BibitemOpen
  \bibfield  {author} {\bibinfo {author} {\bibfnamefont {I.}~\bibnamefont
  {Tews}},\ }\href {\doibase 10.3389/fphy.2020.00153} {\bibfield  {journal}
  {\bibinfo  {journal} {Front. in Phys.}\ }\textbf {\bibinfo {volume} {8}},\
  \bibinfo {pages} {153} (\bibinfo {year} {2020})}\BibitemShut {NoStop}%
\bibitem [{\citenamefont {Hergert}(2020)}]{Hergert:2020bxy}%
  \BibitemOpen
  \bibfield  {author} {\bibinfo {author} {\bibfnamefont {H.}~\bibnamefont
  {Hergert}},\ }\href {\doibase 10.3389/fphy.2020.00379} {\bibfield  {journal}
  {\bibinfo  {journal} {Front. in Phys.}\ }\textbf {\bibinfo {volume} {8}},\
  \bibinfo {pages} {379} (\bibinfo {year} {2020})},\ \Eprint
  {http://arxiv.org/abs/2008.05061} {arXiv:2008.05061} \BibitemShut {NoStop}%
\bibitem [{\citenamefont {Stroberg}\ \emph {et~al.}(2019)\citenamefont
  {Stroberg}, \citenamefont {Bogner}, \citenamefont {Hergert},\ and\
  \citenamefont {Holt}}]{Stroberg:2019mxo}%
  \BibitemOpen
  \bibfield  {author} {\bibinfo {author} {\bibfnamefont {S.~R.}\ \bibnamefont
  {Stroberg}}, \bibinfo {author} {\bibfnamefont {S.~K.}\ \bibnamefont
  {Bogner}}, \bibinfo {author} {\bibfnamefont {H.}~\bibnamefont {Hergert}}, \
  and\ \bibinfo {author} {\bibfnamefont {J.~D.}\ \bibnamefont {Holt}},\ }\href
  {\doibase 10.1146/annurev-nucl-101917-021120} {\bibfield  {journal} {\bibinfo
   {journal} {Annu. Rev. Nucl. Part. Sci.}\ }\textbf {\bibinfo {volume} {69}},\
  \bibinfo {pages} {307} (\bibinfo {year} {2019})},\ \Eprint
  {http://arxiv.org/abs/1902.06154} {arXiv:1902.06154} \BibitemShut {NoStop}%
\bibitem [{\citenamefont {Sivia}\ and\ \citenamefont
  {Skilling}(2006)}]{Sivia:2006}%
  \BibitemOpen
  \bibfield  {author} {\bibinfo {author} {\bibfnamefont {D.}~\bibnamefont
  {Sivia}}\ and\ \bibinfo {author} {\bibfnamefont {J.}~\bibnamefont
  {Skilling}},\ }\href@noop {} {\emph {\bibinfo {title} {Data Analysis: A
  Bayesian Tutorial}}}\ (\bibinfo  {publisher} {Oxford University Press},\
  \bibinfo {year} {2006})\BibitemShut {NoStop}%
\bibitem [{\citenamefont {Melendez}(2020)}]{Melendez:2020xcs}%
  \BibitemOpen
  \bibfield  {author} {\bibinfo {author} {\bibfnamefont {J.}~\bibnamefont
  {Melendez}},\ }\emph {\bibinfo {title} {{Effective Field Theory Truncation
  Errors and Why They Matter}}},\ \href
  {http://rave.ohiolink.edu/etdc/view?acc_num=osu1587114253866152} {Ph.D.
  thesis},\ \bibinfo  {school} {Ohio State U.} (\bibinfo {year}
  {2020})\BibitemShut {NoStop}%
\bibitem [{\citenamefont {Phillips}\ \emph {et~al.}(2021)\citenamefont
  {Phillips}, \citenamefont {Furnstahl}, \citenamefont {Heinz}, \citenamefont
  {Maiti}, \citenamefont {Nazarewicz}, \citenamefont {Nunes}, \citenamefont
  {Plumlee}, \citenamefont {Pratola}, \citenamefont {Pratt}, \citenamefont
  {Viens},\ and\ \citenamefont {Wild}}]{Phillips:2020dmw}%
  \BibitemOpen
  \bibfield  {author} {\bibinfo {author} {\bibfnamefont {D.~R.}\ \bibnamefont
  {Phillips}}, \bibinfo {author} {\bibfnamefont {R.~J.}\ \bibnamefont
  {Furnstahl}}, \bibinfo {author} {\bibfnamefont {U.}~\bibnamefont {Heinz}},
  \bibinfo {author} {\bibfnamefont {T.}~\bibnamefont {Maiti}}, \bibinfo
  {author} {\bibfnamefont {W.}~\bibnamefont {Nazarewicz}}, \bibinfo {author}
  {\bibfnamefont {F.~M.}\ \bibnamefont {Nunes}}, \bibinfo {author}
  {\bibfnamefont {M.}~\bibnamefont {Plumlee}}, \bibinfo {author} {\bibfnamefont
  {M.~T.}\ \bibnamefont {Pratola}}, \bibinfo {author} {\bibfnamefont
  {S.}~\bibnamefont {Pratt}}, \bibinfo {author} {\bibfnamefont {F.~G.}\
  \bibnamefont {Viens}}, \ and\ \bibinfo {author} {\bibfnamefont {S.~M.}\
  \bibnamefont {Wild}},\ }\href {\doibase 10.1088/1361-6471/abf1df} {\bibfield
  {journal} {\bibinfo  {journal} {J. Phys. G}\ }\textbf {\bibinfo {volume}
  {48}},\ \bibinfo {pages} {072001} (\bibinfo {year} {2021})},\ \Eprint
  {http://arxiv.org/abs/2012.07704} {arXiv:2012.07704 [nucl-th]} \BibitemShut
  {NoStop}%
\bibitem [{\citenamefont {Melendez}\ \emph {et~al.}(2019)\citenamefont
  {Melendez}, \citenamefont {Furnstahl}, \citenamefont {Phillips},
  \citenamefont {Pratola},\ and\ \citenamefont
  {Wesolowski}}]{Melendez:2019izc}%
  \BibitemOpen
  \bibfield  {author} {\bibinfo {author} {\bibfnamefont {J.~A.}\ \bibnamefont
  {Melendez}}, \bibinfo {author} {\bibfnamefont {R.~J.}\ \bibnamefont
  {Furnstahl}}, \bibinfo {author} {\bibfnamefont {D.~R.}\ \bibnamefont
  {Phillips}}, \bibinfo {author} {\bibfnamefont {M.~T.}\ \bibnamefont
  {Pratola}}, \ and\ \bibinfo {author} {\bibfnamefont {S.}~\bibnamefont
  {Wesolowski}},\ }\href {\doibase 10.1103/PhysRevC.100.044001} {\bibfield
  {journal} {\bibinfo  {journal} {Phys. Rev. C}\ }\textbf {\bibinfo {volume}
  {100}},\ \bibinfo {pages} {044001} (\bibinfo {year} {2019})},\ \Eprint
  {http://arxiv.org/abs/1904.10581} {arXiv:1904.10581} \BibitemShut {NoStop}%
\bibitem [{\citenamefont {Furnstahl}\ \emph {et~al.}(2015)\citenamefont
  {Furnstahl}, \citenamefont {Phillips},\ and\ \citenamefont
  {Wesolowski}}]{Furnstahl:2014xsa}%
  \BibitemOpen
  \bibfield  {author} {\bibinfo {author} {\bibfnamefont {R.~J.}\ \bibnamefont
  {Furnstahl}}, \bibinfo {author} {\bibfnamefont {D.~R.}\ \bibnamefont
  {Phillips}}, \ and\ \bibinfo {author} {\bibfnamefont {S.}~\bibnamefont
  {Wesolowski}},\ }\href {\doibase 10.1088/0954-3899/42/3/034028} {\bibfield
  {journal} {\bibinfo  {journal} {J. Phys. G}\ }\textbf {\bibinfo {volume}
  {42}},\ \bibinfo {pages} {034028} (\bibinfo {year} {2015})},\ \Eprint
  {http://arxiv.org/abs/1407.0657} {arXiv:1407.0657} \BibitemShut {NoStop}%
\bibitem [{\citenamefont {Sarkar}\ and\ \citenamefont
  {Lee}(2021)}]{Sarkar:2021fpz}%
  \BibitemOpen
  \bibfield  {author} {\bibinfo {author} {\bibfnamefont {A.}~\bibnamefont
  {Sarkar}}\ and\ \bibinfo {author} {\bibfnamefont {D.}~\bibnamefont {Lee}}\
  }(\bibinfo {year} {2021})\ \Eprint {http://arxiv.org/abs/2107.13449}
  {arXiv:2107.13449 [nucl-th]} \BibitemShut {NoStop}%
\bibitem [{\citenamefont {Frame}\ \emph {et~al.}(2018)\citenamefont {Frame},
  \citenamefont {He}, \citenamefont {Ipsen}, \citenamefont {Lee}, \citenamefont
  {Lee},\ and\ \citenamefont {Rrapaj}}]{Frame:2017fah}%
  \BibitemOpen
  \bibfield  {author} {\bibinfo {author} {\bibfnamefont {D.}~\bibnamefont
  {Frame}}, \bibinfo {author} {\bibfnamefont {R.}~\bibnamefont {He}}, \bibinfo
  {author} {\bibfnamefont {I.}~\bibnamefont {Ipsen}}, \bibinfo {author}
  {\bibfnamefont {D.}~\bibnamefont {Lee}}, \bibinfo {author} {\bibfnamefont
  {D.}~\bibnamefont {Lee}}, \ and\ \bibinfo {author} {\bibfnamefont
  {E.}~\bibnamefont {Rrapaj}},\ }\href {\doibase
  10.1103/PhysRevLett.121.032501} {\bibfield  {journal} {\bibinfo  {journal}
  {Phys. Rev. Lett.}\ }\textbf {\bibinfo {volume} {121}},\ \bibinfo {pages}
  {032501} (\bibinfo {year} {2018})},\ \Eprint
  {http://arxiv.org/abs/1711.07090} {arXiv:1711.07090} \BibitemShut {NoStop}%
\bibitem [{\citenamefont {Melendez}\ \emph {et~al.}(2021)\citenamefont
  {Melendez}, \citenamefont {Drischler}, \citenamefont {Garcia}, \citenamefont
  {Furnstahl},\ and\ \citenamefont {Zhang}}]{Melendez:2021lyq}%
  \BibitemOpen
  \bibfield  {author} {\bibinfo {author} {\bibfnamefont {J.}~\bibnamefont
  {Melendez}}, \bibinfo {author} {\bibfnamefont {C.}~\bibnamefont {Drischler}},
  \bibinfo {author} {\bibfnamefont {A.}~\bibnamefont {Garcia}}, \bibinfo
  {author} {\bibfnamefont {R.}~\bibnamefont {Furnstahl}}, \ and\ \bibinfo
  {author} {\bibfnamefont {X.}~\bibnamefont {Zhang}},\ }\href {\doibase
  https://doi.org/10.1016/j.physletb.2021.136608} {\bibfield  {journal}
  {\bibinfo  {journal} {Phys. Lett. B}\ }\textbf {\bibinfo {volume} {821}},\
  \bibinfo {pages} {136608} (\bibinfo {year} {2021})}\BibitemShut {NoStop}%
\bibitem [{\citenamefont {Wesolowski}\ \emph {et~al.}(2021)\citenamefont
  {Wesolowski}, \citenamefont {Svensson}, \citenamefont {Ekstr\"om},
  \citenamefont {Forss\'en}, \citenamefont {Furnstahl}, \citenamefont
  {Melendez},\ and\ \citenamefont {Phillips}}]{Wesolowski:2021cni}%
  \BibitemOpen
  \bibfield  {author} {\bibinfo {author} {\bibfnamefont {S.}~\bibnamefont
  {Wesolowski}}, \bibinfo {author} {\bibfnamefont {I.}~\bibnamefont
  {Svensson}}, \bibinfo {author} {\bibfnamefont {A.}~\bibnamefont {Ekstr\"om}},
  \bibinfo {author} {\bibfnamefont {C.}~\bibnamefont {Forss\'en}}, \bibinfo
  {author} {\bibfnamefont {R.~J.}\ \bibnamefont {Furnstahl}}, \bibinfo {author}
  {\bibfnamefont {J.~A.}\ \bibnamefont {Melendez}}, \ and\ \bibinfo {author}
  {\bibfnamefont {D.~R.}\ \bibnamefont {Phillips}},\ }\href@noop {} {\
  (\bibinfo {year} {2021})},\ \Eprint {http://arxiv.org/abs/2104.04441}
  {arXiv:2104.04441 [nucl-th]} \BibitemShut {NoStop}%
\bibitem [{\citenamefont {Miller}\ \emph
  {et~al.}(2021{\natexlab{a}})\citenamefont {Miller}, \citenamefont
  {Ekstr\"om},\ and\ \citenamefont {Forss\'en}}]{Miller:2021pcu}%
  \BibitemOpen
  \bibfield  {author} {\bibinfo {author} {\bibfnamefont {S.~B.~S.}\
  \bibnamefont {Miller}}, \bibinfo {author} {\bibfnamefont {A.}~\bibnamefont
  {Ekstr\"om}}, \ and\ \bibinfo {author} {\bibfnamefont {C.}~\bibnamefont
  {Forss\'en}},\ }\href@noop {} {\  (\bibinfo {year} {2021}{\natexlab{a}})},\
  \Eprint {http://arxiv.org/abs/2106.00454} {arXiv:2106.00454 [nucl-th]}
  \BibitemShut {NoStop}%
\bibitem [{\citenamefont {Furnstahl}\ \emph {et~al.}(2020)\citenamefont
  {Furnstahl}, \citenamefont {Garcia}, \citenamefont {Millican},\ and\
  \citenamefont {Zhang}}]{Furnstahl:2020abp}%
  \BibitemOpen
  \bibfield  {author} {\bibinfo {author} {\bibfnamefont {R.~J.}\ \bibnamefont
  {Furnstahl}}, \bibinfo {author} {\bibfnamefont {A.~J.}\ \bibnamefont
  {Garcia}}, \bibinfo {author} {\bibfnamefont {P.~J.}\ \bibnamefont
  {Millican}}, \ and\ \bibinfo {author} {\bibfnamefont {X.}~\bibnamefont
  {Zhang}},\ }\href {\doibase 10.1016/j.physletb.2020.135719} {\bibfield
  {journal} {\bibinfo  {journal} {Phys. Lett. B}\ }\textbf {\bibinfo {volume}
  {809}},\ \bibinfo {pages} {135719} (\bibinfo {year} {2020})},\ \Eprint
  {http://arxiv.org/abs/2007.03635} {arXiv:2007.03635 [nucl-th]} \BibitemShut
  {NoStop}%
\bibitem [{\citenamefont {Ekström}\ and\ \citenamefont
  {Hagen}(2019)}]{Ekstrom:2019lss}%
  \BibitemOpen
  \bibfield  {author} {\bibinfo {author} {\bibfnamefont {A.}~\bibnamefont
  {Ekström}}\ and\ \bibinfo {author} {\bibfnamefont {G.}~\bibnamefont
  {Hagen}},\ }\href {\doibase 10.1103/PhysRevLett.123.252501} {\bibfield
  {journal} {\bibinfo  {journal} {Phys. Rev. Lett.}\ }\textbf {\bibinfo
  {volume} {123}},\ \bibinfo {pages} {252501} (\bibinfo {year} {2019})},\
  \Eprint {http://arxiv.org/abs/1910.02922} {arXiv:1910.02922 [nucl-th]}
  \BibitemShut {NoStop}%
\bibitem [{\citenamefont {K\"onig}\ \emph {et~al.}(2020)\citenamefont
  {K\"onig}, \citenamefont {Ekstr\"om}, \citenamefont {Hebeler}, \citenamefont
  {Lee},\ and\ \citenamefont {Schwenk}}]{Konig:2019adq}%
  \BibitemOpen
  \bibfield  {author} {\bibinfo {author} {\bibfnamefont {S.}~\bibnamefont
  {K\"onig}}, \bibinfo {author} {\bibfnamefont {A.}~\bibnamefont {Ekstr\"om}},
  \bibinfo {author} {\bibfnamefont {K.}~\bibnamefont {Hebeler}}, \bibinfo
  {author} {\bibfnamefont {D.}~\bibnamefont {Lee}}, \ and\ \bibinfo {author}
  {\bibfnamefont {A.}~\bibnamefont {Schwenk}},\ }\href {\doibase
  10.1016/j.physletb.2020.135814} {\bibfield  {journal} {\bibinfo  {journal}
  {Phys. Lett. B}\ }\textbf {\bibinfo {volume} {810}},\ \bibinfo {pages}
  {135814} (\bibinfo {year} {2020})},\ \Eprint
  {http://arxiv.org/abs/1909.08446} {arXiv:1909.08446 [nucl-th]} \BibitemShut
  {NoStop}%
\bibitem [{\citenamefont {Jiang}\ \emph {et~al.}(2020)\citenamefont {Jiang},
  \citenamefont {Ekstr\"om}, \citenamefont {Forss\'en}, \citenamefont {Hagen},
  \citenamefont {Jansen},\ and\ \citenamefont {Papenbrock}}]{Jiang:2020the}%
  \BibitemOpen
  \bibfield  {author} {\bibinfo {author} {\bibfnamefont {W.}~\bibnamefont
  {Jiang}}, \bibinfo {author} {\bibfnamefont {A.}~\bibnamefont {Ekstr\"om}},
  \bibinfo {author} {\bibfnamefont {C.}~\bibnamefont {Forss\'en}}, \bibinfo
  {author} {\bibfnamefont {G.}~\bibnamefont {Hagen}}, \bibinfo {author}
  {\bibfnamefont {G.}~\bibnamefont {Jansen}}, \ and\ \bibinfo {author}
  {\bibfnamefont {T.}~\bibnamefont {Papenbrock}},\ }\href {\doibase
  10.1103/PhysRevC.102.054301} {\bibfield  {journal} {\bibinfo  {journal}
  {Phys. Rev. C}\ }\textbf {\bibinfo {volume} {102}},\ \bibinfo {pages}
  {054301} (\bibinfo {year} {2020})},\ \Eprint
  {http://arxiv.org/abs/2006.16774} {arXiv:2006.16774} \BibitemShut {NoStop}%
\bibitem [{\citenamefont {Piarulli}\ and\ \citenamefont
  {Tews}(2020)}]{Piarulli:2019cqu}%
  \BibitemOpen
  \bibfield  {author} {\bibinfo {author} {\bibfnamefont {M.}~\bibnamefont
  {Piarulli}}\ and\ \bibinfo {author} {\bibfnamefont {I.}~\bibnamefont
  {Tews}},\ }\href {\doibase 10.3389/fphy.2019.00245} {\bibfield  {journal}
  {\bibinfo  {journal} {Front. in Phys.}\ }\textbf {\bibinfo {volume} {7}},\
  \bibinfo {pages} {245} (\bibinfo {year} {2020})},\ \Eprint
  {http://arxiv.org/abs/2002.00032} {arXiv:2002.00032 [nucl-th]} \BibitemShut
  {NoStop}%
\bibitem [{\citenamefont {Ekström}\ \emph {et~al.}(2018)\citenamefont
  {Ekström}, \citenamefont {Hagen}, \citenamefont {Morris}, \citenamefont
  {Papenbrock},\ and\ \citenamefont {Schwartz}}]{Ekstrom:2017koy}%
  \BibitemOpen
  \bibfield  {author} {\bibinfo {author} {\bibfnamefont {A.}~\bibnamefont
  {Ekström}}, \bibinfo {author} {\bibfnamefont {G.}~\bibnamefont {Hagen}},
  \bibinfo {author} {\bibfnamefont {T.~D.}\ \bibnamefont {Morris}}, \bibinfo
  {author} {\bibfnamefont {T.}~\bibnamefont {Papenbrock}}, \ and\ \bibinfo
  {author} {\bibfnamefont {P.~D.}\ \bibnamefont {Schwartz}},\ }\href {\doibase
  10.1103/PhysRevC.97.024332} {\bibfield  {journal} {\bibinfo  {journal} {Phys.
  Rev. C}\ }\textbf {\bibinfo {volume} {97}},\ \bibinfo {pages} {024332}
  (\bibinfo {year} {2018})},\ \Eprint {http://arxiv.org/abs/1707.09028}
  {arXiv:1707.09028} \BibitemShut {NoStop}%
\bibitem [{\citenamefont {Drischler}\ \emph
  {et~al.}(2021{\natexlab{b}})\citenamefont {Drischler}, \citenamefont
  {Haxton}, \citenamefont {McElvain}, \citenamefont {Mereghetti}, \citenamefont
  {Nicholson}, \citenamefont {Vranas},\ and\ \citenamefont
  {Walker-Loud}}]{DRISCHLER2021103888}%
  \BibitemOpen
  \bibfield  {author} {\bibinfo {author} {\bibfnamefont {C.}~\bibnamefont
  {Drischler}}, \bibinfo {author} {\bibfnamefont {W.}~\bibnamefont {Haxton}},
  \bibinfo {author} {\bibfnamefont {K.}~\bibnamefont {McElvain}}, \bibinfo
  {author} {\bibfnamefont {E.}~\bibnamefont {Mereghetti}}, \bibinfo {author}
  {\bibfnamefont {A.}~\bibnamefont {Nicholson}}, \bibinfo {author}
  {\bibfnamefont {P.}~\bibnamefont {Vranas}}, \ and\ \bibinfo {author}
  {\bibfnamefont {A.}~\bibnamefont {Walker-Loud}},\ }\href {\doibase
  https://doi.org/10.1016/j.ppnp.2021.103888} {\bibfield  {journal} {\bibinfo
  {journal} {Prog. Part. Nucl. Phys.}\ }\textbf {\bibinfo {volume} {XY}},\
  \bibinfo {pages} {103888} (\bibinfo {year} {2021}{\natexlab{b}})},\ \bibinfo
  {note} {in press}\BibitemShut {NoStop}%
\bibitem [{\citenamefont {{{Low Energy Nuclear Physics International
  Collaboration (LENPIC)}}}(2021)}]{LENPICweb}%
  \BibitemOpen
  \bibfield  {author} {\bibinfo {author} {\bibnamefont {{{Low Energy Nuclear
  Physics International Collaboration (LENPIC)}}}}\ }(\bibinfo {year} {2021})\
  \bibinfo {note} {\url{http://www.lenpic.org/}}\BibitemShut {NoStop}%
\bibitem [{\citenamefont {Hergert}\ \emph {et~al.}(2016)\citenamefont
  {Hergert}, \citenamefont {Bogner}, \citenamefont {Morris}, \citenamefont
  {Schwenk},\ and\ \citenamefont {Tsukiyama}}]{Hergert:2015awm}%
  \BibitemOpen
  \bibfield  {author} {\bibinfo {author} {\bibfnamefont {H.}~\bibnamefont
  {Hergert}}, \bibinfo {author} {\bibfnamefont {S.~K.}\ \bibnamefont {Bogner}},
  \bibinfo {author} {\bibfnamefont {T.~D.}\ \bibnamefont {Morris}}, \bibinfo
  {author} {\bibfnamefont {A.}~\bibnamefont {Schwenk}}, \ and\ \bibinfo
  {author} {\bibfnamefont {K.}~\bibnamefont {Tsukiyama}},\ }\href {\doibase
  10.1016/j.physrep.2015.12.007} {\bibfield  {journal} {\bibinfo  {journal}
  {Phys. Rept.}\ }\textbf {\bibinfo {volume} {621}},\ \bibinfo {pages} {165}
  (\bibinfo {year} {2016})},\ \Eprint {http://arxiv.org/abs/1512.06956}
  {arXiv:1512.06956} \BibitemShut {NoStop}%
\bibitem [{\citenamefont {Morris}(2016)}]{Titusthesis}%
  \BibitemOpen
  \bibfield  {author} {\bibinfo {author} {\bibfnamefont {T.}~\bibnamefont
  {Morris}},\ }\emph {\bibinfo {title} {Systematic improvements of ab-initio
  in-medium similarity renormalization group calculations}},\ \href {\doibase
  10.25335/M51W9R} {Ph.D. thesis},\ \bibinfo  {school} {Michigan State
  University} (\bibinfo {year} {2016})\BibitemShut {NoStop}%
\bibitem [{\citenamefont {Heinz}\ \emph {et~al.}(2021)\citenamefont {Heinz},
  \citenamefont {Tichai}, \citenamefont {Hoppe}, \citenamefont {Hebeler},\ and\
  \citenamefont {Schwenk}}]{Heinz:2021xir}%
  \BibitemOpen
  \bibfield  {author} {\bibinfo {author} {\bibfnamefont {M.}~\bibnamefont
  {Heinz}}, \bibinfo {author} {\bibfnamefont {A.}~\bibnamefont {Tichai}},
  \bibinfo {author} {\bibfnamefont {J.}~\bibnamefont {Hoppe}}, \bibinfo
  {author} {\bibfnamefont {K.}~\bibnamefont {Hebeler}}, \ and\ \bibinfo
  {author} {\bibfnamefont {A.}~\bibnamefont {Schwenk}},\ }\href {\doibase
  10.1103/PhysRevC.103.044318} {\bibfield  {journal} {\bibinfo  {journal}
  {Phys. Rev. C}\ }\textbf {\bibinfo {volume} {103}},\ \bibinfo {pages}
  {044318} (\bibinfo {year} {2021})},\ \Eprint
  {http://arxiv.org/abs/2102.11172} {arXiv:2102.11172 [nucl-th]} \BibitemShut
  {NoStop}%
\bibitem [{\citenamefont {Weinberg}(1963)}]{PhysRev.131.440}%
  \BibitemOpen
  \bibfield  {author} {\bibinfo {author} {\bibfnamefont {S.}~\bibnamefont
  {Weinberg}},\ }\href {\doibase 10.1103/PhysRev.131.440} {\bibfield  {journal}
  {\bibinfo  {journal} {Phys. Rev.}\ }\textbf {\bibinfo {volume} {131}},\
  \bibinfo {pages} {440} (\bibinfo {year} {1963})}\BibitemShut {NoStop}%
\bibitem [{\citenamefont {Hoppe}\ \emph {et~al.}(2017)\citenamefont {Hoppe},
  \citenamefont {Drischler}, \citenamefont {Furnstahl}, \citenamefont
  {Hebeler},\ and\ \citenamefont {Schwenk}}]{Hoppe:2017lok}%
  \BibitemOpen
  \bibfield  {author} {\bibinfo {author} {\bibfnamefont {J.}~\bibnamefont
  {Hoppe}}, \bibinfo {author} {\bibfnamefont {C.}~\bibnamefont {Drischler}},
  \bibinfo {author} {\bibfnamefont {R.~J.}\ \bibnamefont {Furnstahl}}, \bibinfo
  {author} {\bibfnamefont {K.}~\bibnamefont {Hebeler}}, \ and\ \bibinfo
  {author} {\bibfnamefont {A.}~\bibnamefont {Schwenk}},\ }\href {\doibase
  10.1103/PhysRevC.96.054002} {\bibfield  {journal} {\bibinfo  {journal} {Phys.
  Rev. C}\ }\textbf {\bibinfo {volume} {96}},\ \bibinfo {pages} {054002}
  (\bibinfo {year} {2017})},\ \Eprint {http://arxiv.org/abs/1707.06438}
  {arXiv:1707.06438} \BibitemShut {NoStop}%
\bibitem [{\citenamefont {Ramanan}\ \emph {et~al.}(2007)\citenamefont
  {Ramanan}, \citenamefont {Bogner},\ and\ \citenamefont
  {Furnstahl}}]{Ramanan:2007bb}%
  \BibitemOpen
  \bibfield  {author} {\bibinfo {author} {\bibfnamefont {S.}~\bibnamefont
  {Ramanan}}, \bibinfo {author} {\bibfnamefont {S.~K.}\ \bibnamefont {Bogner}},
  \ and\ \bibinfo {author} {\bibfnamefont {R.~J.}\ \bibnamefont {Furnstahl}},\
  }\href {\doibase 10.1016/j.nuclphysa.2007.10.005} {\bibfield  {journal}
  {\bibinfo  {journal} {Nucl. Phys. A}\ }\textbf {\bibinfo {volume} {797}},\
  \bibinfo {pages} {81} (\bibinfo {year} {2007})},\ \Eprint
  {http://arxiv.org/abs/0709.0534} {arXiv:0709.0534} \BibitemShut {NoStop}%
\bibitem [{\citenamefont {Srinivas}\ and\ \citenamefont
  {Ramanan}(2016)}]{Srinivas:2016kir}%
  \BibitemOpen
  \bibfield  {author} {\bibinfo {author} {\bibfnamefont {S.}~\bibnamefont
  {Srinivas}}\ and\ \bibinfo {author} {\bibfnamefont {S.}~\bibnamefont
  {Ramanan}},\ }\href {\doibase 10.1103/PhysRevC.94.064303} {\bibfield
  {journal} {\bibinfo  {journal} {Phys. Rev. C}\ }\textbf {\bibinfo {volume}
  {94}},\ \bibinfo {pages} {064303} (\bibinfo {year} {2016})},\ \Eprint
  {http://arxiv.org/abs/1606.09053} {arXiv:1606.09053 [nucl-th]} \BibitemShut
  {NoStop}%
\bibitem [{\citenamefont {Reinert}\ \emph {et~al.}(2018)\citenamefont
  {Reinert}, \citenamefont {Krebs},\ and\ \citenamefont
  {Epelbaum}}]{Reinert:2017usi}%
  \BibitemOpen
  \bibfield  {author} {\bibinfo {author} {\bibfnamefont {P.}~\bibnamefont
  {Reinert}}, \bibinfo {author} {\bibfnamefont {H.}~\bibnamefont {Krebs}}, \
  and\ \bibinfo {author} {\bibfnamefont {E.}~\bibnamefont {Epelbaum}},\ }\href
  {\doibase 10.1140/epja/i2018-12516-4} {\bibfield  {journal} {\bibinfo
  {journal} {Eur. Phys. J. A}\ }\textbf {\bibinfo {volume} {54}},\ \bibinfo
  {pages} {86} (\bibinfo {year} {2018})},\ \Eprint
  {http://arxiv.org/abs/1711.08821} {arXiv:1711.08821} \BibitemShut {NoStop}%
\bibitem [{\citenamefont {Hagen}\ \emph {et~al.}(2015)\citenamefont {Hagen}
  \emph {et~al.}}]{Hagen:2015yea}%
  \BibitemOpen
  \bibfield  {author} {\bibinfo {author} {\bibfnamefont {G.}~\bibnamefont
  {Hagen}} \emph {et~al.},\ }\href {\doibase 10.1038/nphys3529} {\bibfield
  {journal} {\bibinfo  {journal} {Nature Phys.}\ }\textbf {\bibinfo {volume}
  {12}},\ \bibinfo {pages} {186} (\bibinfo {year} {2015})},\ \Eprint
  {http://arxiv.org/abs/1509.07169} {arXiv:1509.07169} \BibitemShut {NoStop}%
\bibitem [{\citenamefont {van Kolck}(2021)}]{vanKolck:2021rqu}%
  \BibitemOpen
  \bibfield  {author} {\bibinfo {author} {\bibfnamefont {U.}~\bibnamefont {van
  Kolck}}\ }(\bibinfo {year} {2021})\ \Eprint {http://arxiv.org/abs/2107.11675}
  {arXiv:2107.11675} \BibitemShut {NoStop}%
\bibitem [{\citenamefont {Maris}\ \emph {et~al.}(2021)\citenamefont {Maris}
  \emph {et~al.}}]{Maris:2020qne}%
  \BibitemOpen
  \bibfield  {author} {\bibinfo {author} {\bibfnamefont {P.}~\bibnamefont
  {Maris}} \emph {et~al.},\ }\href {\doibase 10.1103/PhysRevC.103.054001}
  {\bibfield  {journal} {\bibinfo  {journal} {Phys. Rev. C}\ }\textbf {\bibinfo
  {volume} {103}},\ \bibinfo {pages} {054001} (\bibinfo {year} {2021})},\
  \Eprint {http://arxiv.org/abs/2012.12396} {arXiv:2012.12396 [nucl-th]}
  \BibitemShut {NoStop}%
\bibitem [{\citenamefont {Bogner}\ \emph {et~al.}(2014)\citenamefont {Bogner},
  \citenamefont {Hergert}, \citenamefont {Holt}, \citenamefont {Schwenk},
  \citenamefont {Binder}, \citenamefont {Calci}, \citenamefont {Langhammer},\
  and\ \citenamefont {Roth}}]{Bogner:2014baa}%
  \BibitemOpen
  \bibfield  {author} {\bibinfo {author} {\bibfnamefont {S.~K.}\ \bibnamefont
  {Bogner}}, \bibinfo {author} {\bibfnamefont {H.}~\bibnamefont {Hergert}},
  \bibinfo {author} {\bibfnamefont {J.~D.}\ \bibnamefont {Holt}}, \bibinfo
  {author} {\bibfnamefont {A.}~\bibnamefont {Schwenk}}, \bibinfo {author}
  {\bibfnamefont {S.}~\bibnamefont {Binder}}, \bibinfo {author} {\bibfnamefont
  {A.}~\bibnamefont {Calci}}, \bibinfo {author} {\bibfnamefont
  {J.}~\bibnamefont {Langhammer}}, \ and\ \bibinfo {author} {\bibfnamefont
  {R.}~\bibnamefont {Roth}},\ }\href {\doibase 10.1103/PhysRevLett.113.142501}
  {\bibfield  {journal} {\bibinfo  {journal} {Phys. Rev. Lett.}\ }\textbf
  {\bibinfo {volume} {113}},\ \bibinfo {pages} {142501} (\bibinfo {year}
  {2014})},\ \Eprint {http://arxiv.org/abs/1402.1407} {arXiv:1402.1407
  [nucl-th]} \BibitemShut {NoStop}%
\bibitem [{\citenamefont {Stroberg}\ \emph {et~al.}(2017)\citenamefont
  {Stroberg}, \citenamefont {Calci}, \citenamefont {Hergert}, \citenamefont
  {Holt}, \citenamefont {Bogner}, \citenamefont {Roth},\ and\ \citenamefont
  {Schwenk}}]{Stroberg:2016ung}%
  \BibitemOpen
  \bibfield  {author} {\bibinfo {author} {\bibfnamefont {S.~R.}\ \bibnamefont
  {Stroberg}}, \bibinfo {author} {\bibfnamefont {A.}~\bibnamefont {Calci}},
  \bibinfo {author} {\bibfnamefont {H.}~\bibnamefont {Hergert}}, \bibinfo
  {author} {\bibfnamefont {J.~D.}\ \bibnamefont {Holt}}, \bibinfo {author}
  {\bibfnamefont {S.~K.}\ \bibnamefont {Bogner}}, \bibinfo {author}
  {\bibfnamefont {R.}~\bibnamefont {Roth}}, \ and\ \bibinfo {author}
  {\bibfnamefont {A.}~\bibnamefont {Schwenk}},\ }\href {\doibase
  10.1103/PhysRevLett.118.032502} {\bibfield  {journal} {\bibinfo  {journal}
  {Phys. Rev. Lett.}\ }\textbf {\bibinfo {volume} {118}},\ \bibinfo {pages}
  {032502} (\bibinfo {year} {2017})},\ \Eprint
  {http://arxiv.org/abs/1607.03229} {arXiv:1607.03229 [nucl-th]} \BibitemShut
  {NoStop}%
\bibitem [{\citenamefont {Sun}\ \emph {et~al.}(2021)\citenamefont {Sun},
  \citenamefont {Hagen}, \citenamefont {Jansen},\ and\ \citenamefont
  {Papenbrock}}]{Sun:2021ffi}%
  \BibitemOpen
  \bibfield  {author} {\bibinfo {author} {\bibfnamefont {Z.~H.}\ \bibnamefont
  {Sun}}, \bibinfo {author} {\bibfnamefont {G.}~\bibnamefont {Hagen}}, \bibinfo
  {author} {\bibfnamefont {G.~R.}\ \bibnamefont {Jansen}}, \ and\ \bibinfo
  {author} {\bibfnamefont {T.}~\bibnamefont {Papenbrock}},\ }\href@noop {} {\
  (\bibinfo {year} {2021})},\ \Eprint {http://arxiv.org/abs/2107.14314}
  {arXiv:2107.14314 [nucl-th]} \BibitemShut {NoStop}%
\bibitem [{\citenamefont {Stroberg}\ \emph {et~al.}(2021)\citenamefont
  {Stroberg}, \citenamefont {Holt}, \citenamefont {Schwenk},\ and\
  \citenamefont {Simonis}}]{Stroberg:2019gmc}%
  \BibitemOpen
  \bibfield  {author} {\bibinfo {author} {\bibfnamefont {S.}~\bibnamefont
  {Stroberg}}, \bibinfo {author} {\bibfnamefont {J.}~\bibnamefont {Holt}},
  \bibinfo {author} {\bibfnamefont {A.}~\bibnamefont {Schwenk}}, \ and\
  \bibinfo {author} {\bibfnamefont {J.}~\bibnamefont {Simonis}},\ }\href
  {\doibase 10.1103/PhysRevLett.126.022501} {\bibfield  {journal} {\bibinfo
  {journal} {Phys. Rev. Lett.}\ }\textbf {\bibinfo {volume} {126}},\ \bibinfo
  {pages} {022501} (\bibinfo {year} {2021})},\ \Eprint
  {http://arxiv.org/abs/1905.10475} {arXiv:1905.10475} \BibitemShut {NoStop}%
\bibitem [{\citenamefont {Hebeler}\ \emph {et~al.}(2011)\citenamefont
  {Hebeler}, \citenamefont {Bogner}, \citenamefont {Furnstahl}, \citenamefont
  {Nogga},\ and\ \citenamefont {Schwenk}}]{Hebeler:2010xb}%
  \BibitemOpen
  \bibfield  {author} {\bibinfo {author} {\bibfnamefont {K.}~\bibnamefont
  {Hebeler}}, \bibinfo {author} {\bibfnamefont {S.~K.}\ \bibnamefont {Bogner}},
  \bibinfo {author} {\bibfnamefont {R.~J.}\ \bibnamefont {Furnstahl}}, \bibinfo
  {author} {\bibfnamefont {A.}~\bibnamefont {Nogga}}, \ and\ \bibinfo {author}
  {\bibfnamefont {A.}~\bibnamefont {Schwenk}},\ }\href {\doibase
  10.1103/PhysRevC.83.031301} {\bibfield  {journal} {\bibinfo  {journal} {Phys.
  Rev. C}\ }\textbf {\bibinfo {volume} {83}},\ \bibinfo {pages} {031301(R)}
  (\bibinfo {year} {2011})},\ \Eprint {http://arxiv.org/abs/1012.3381}
  {arXiv:1012.3381} \BibitemShut {NoStop}%
\bibitem [{\citenamefont {Drischler}\ \emph {et~al.}(2019)\citenamefont
  {Drischler}, \citenamefont {Hebeler},\ and\ \citenamefont
  {Schwenk}}]{Drischler:2017wtt}%
  \BibitemOpen
  \bibfield  {author} {\bibinfo {author} {\bibfnamefont {C.}~\bibnamefont
  {Drischler}}, \bibinfo {author} {\bibfnamefont {K.}~\bibnamefont {Hebeler}},
  \ and\ \bibinfo {author} {\bibfnamefont {A.}~\bibnamefont {Schwenk}},\ }\href
  {\doibase 10.1103/PhysRevLett.122.042501} {\bibfield  {journal} {\bibinfo
  {journal} {Phys. Rev. Lett.}\ }\textbf {\bibinfo {volume} {122}},\ \bibinfo
  {pages} {042501} (\bibinfo {year} {2019})},\ \Eprint
  {http://arxiv.org/abs/1710.08220} {arXiv:1710.08220} \BibitemShut {NoStop}%
\bibitem [{\citenamefont {Simonis}\ \emph {et~al.}(2017)\citenamefont
  {Simonis}, \citenamefont {Stroberg}, \citenamefont {Hebeler}, \citenamefont
  {Holt},\ and\ \citenamefont {Schwenk}}]{Simonis:2017dny}%
  \BibitemOpen
  \bibfield  {author} {\bibinfo {author} {\bibfnamefont {J.}~\bibnamefont
  {Simonis}}, \bibinfo {author} {\bibfnamefont {S.}~\bibnamefont {Stroberg}},
  \bibinfo {author} {\bibfnamefont {K.}~\bibnamefont {Hebeler}}, \bibinfo
  {author} {\bibfnamefont {J.}~\bibnamefont {Holt}}, \ and\ \bibinfo {author}
  {\bibfnamefont {A.}~\bibnamefont {Schwenk}},\ }\href {\doibase
  10.1103/PhysRevC.96.014303} {\bibfield  {journal} {\bibinfo  {journal} {Phys.
  Rev. C}\ }\textbf {\bibinfo {volume} {96}},\ \bibinfo {pages} {014303}
  (\bibinfo {year} {2017})},\ \Eprint {http://arxiv.org/abs/1704.02915}
  {arXiv:1704.02915} \BibitemShut {NoStop}%
\bibitem [{\citenamefont {Gysbers}\ \emph {et~al.}(2019)\citenamefont {Gysbers}
  \emph {et~al.}}]{Gysbers:2019uyb}%
  \BibitemOpen
  \bibfield  {author} {\bibinfo {author} {\bibfnamefont {P.}~\bibnamefont
  {Gysbers}} \emph {et~al.},\ }\href {\doibase 10.1038/s41567-019-0450-7}
  {\bibfield  {journal} {\bibinfo  {journal} {Nature Phys.}\ }\textbf {\bibinfo
  {volume} {15}},\ \bibinfo {pages} {428} (\bibinfo {year} {2019})},\ \Eprint
  {http://arxiv.org/abs/1903.00047} {arXiv:1903.00047 [nucl-th]} \BibitemShut
  {NoStop}%
\bibitem [{\citenamefont {Yao}\ \emph {et~al.}(2020)\citenamefont {Yao},
  \citenamefont {Bally}, \citenamefont {Engel}, \citenamefont {Wirth},
  \citenamefont {Rodr\'\i{}guez},\ and\ \citenamefont {Hergert}}]{Yao:2019rck}%
  \BibitemOpen
  \bibfield  {author} {\bibinfo {author} {\bibfnamefont {J.~M.}\ \bibnamefont
  {Yao}}, \bibinfo {author} {\bibfnamefont {B.}~\bibnamefont {Bally}}, \bibinfo
  {author} {\bibfnamefont {J.}~\bibnamefont {Engel}}, \bibinfo {author}
  {\bibfnamefont {R.}~\bibnamefont {Wirth}}, \bibinfo {author} {\bibfnamefont
  {T.~R.}\ \bibnamefont {Rodr\'\i{}guez}}, \ and\ \bibinfo {author}
  {\bibfnamefont {H.}~\bibnamefont {Hergert}},\ }\href {\doibase
  10.1103/PhysRevLett.124.232501} {\bibfield  {journal} {\bibinfo  {journal}
  {Phys. Rev. Lett.}\ }\textbf {\bibinfo {volume} {124}},\ \bibinfo {pages}
  {232501} (\bibinfo {year} {2020})},\ \Eprint
  {http://arxiv.org/abs/1908.05424} {arXiv:1908.05424 [nucl-th]} \BibitemShut
  {NoStop}%
\bibitem [{\citenamefont {Belley}\ \emph {et~al.}(2021)\citenamefont {Belley},
  \citenamefont {Payne}, \citenamefont {Stroberg}, \citenamefont {Miyagi},\
  and\ \citenamefont {Holt}}]{Belley:2020ejd}%
  \BibitemOpen
  \bibfield  {author} {\bibinfo {author} {\bibfnamefont {A.}~\bibnamefont
  {Belley}}, \bibinfo {author} {\bibfnamefont {C.~G.}\ \bibnamefont {Payne}},
  \bibinfo {author} {\bibfnamefont {S.~R.}\ \bibnamefont {Stroberg}}, \bibinfo
  {author} {\bibfnamefont {T.}~\bibnamefont {Miyagi}}, \ and\ \bibinfo {author}
  {\bibfnamefont {J.~D.}\ \bibnamefont {Holt}},\ }\href {\doibase
  10.1103/PhysRevLett.126.042502} {\bibfield  {journal} {\bibinfo  {journal}
  {Phys. Rev. Lett.}\ }\textbf {\bibinfo {volume} {126}},\ \bibinfo {pages}
  {042502} (\bibinfo {year} {2021})},\ \Eprint
  {http://arxiv.org/abs/2008.06588} {arXiv:2008.06588 [nucl-th]} \BibitemShut
  {NoStop}%
\bibitem [{\citenamefont {Epelbaum}\ \emph
  {et~al.}(2015{\natexlab{a}})\citenamefont {Epelbaum}, \citenamefont {Krebs},\
  and\ \citenamefont {Mei{\ss}ner}}]{Epelbaum:2014efa}%
  \BibitemOpen
  \bibfield  {author} {\bibinfo {author} {\bibfnamefont {E.}~\bibnamefont
  {Epelbaum}}, \bibinfo {author} {\bibfnamefont {H.}~\bibnamefont {Krebs}}, \
  and\ \bibinfo {author} {\bibfnamefont {U.~G.}\ \bibnamefont {Mei{\ss}ner}},\
  }\href {\doibase 10.1140/epja/i2015-15053-8} {\bibfield  {journal} {\bibinfo
  {journal} {Eur. Phys. J. A}\ }\textbf {\bibinfo {volume} {51}},\ \bibinfo
  {pages} {53} (\bibinfo {year} {2015}{\natexlab{a}})},\ \Eprint
  {http://arxiv.org/abs/1412.0142} {arXiv:1412.0142} \BibitemShut {NoStop}%
\bibitem [{\citenamefont {Epelbaum}\ \emph
  {et~al.}(2015{\natexlab{b}})\citenamefont {Epelbaum}, \citenamefont {Krebs},\
  and\ \citenamefont {Mei{\ss}ner}}]{Epelbaum:2014sza}%
  \BibitemOpen
  \bibfield  {author} {\bibinfo {author} {\bibfnamefont {E.}~\bibnamefont
  {Epelbaum}}, \bibinfo {author} {\bibfnamefont {H.}~\bibnamefont {Krebs}}, \
  and\ \bibinfo {author} {\bibfnamefont {U.~G.}\ \bibnamefont {Mei{\ss}ner}},\
  }\href {\doibase 10.1103/PhysRevLett.115.122301} {\bibfield  {journal}
  {\bibinfo  {journal} {Phys. Rev. Lett.}\ }\textbf {\bibinfo {volume} {115}},\
  \bibinfo {pages} {122301} (\bibinfo {year} {2015}{\natexlab{b}})},\ \Eprint
  {http://arxiv.org/abs/1412.4623} {arXiv:1412.4623} \BibitemShut {NoStop}%
\bibitem [{\citenamefont {Binder}\ \emph {et~al.}(2016)\citenamefont {Binder},
  \citenamefont {Calci}, \citenamefont {Epelbaum}, \citenamefont {Furnstahl},
  \citenamefont {Golak} \emph {et~al.}}]{Binder:2015mbz}%
  \BibitemOpen
  \bibfield  {author} {\bibinfo {author} {\bibfnamefont {S.}~\bibnamefont
  {Binder}}, \bibinfo {author} {\bibfnamefont {A.}~\bibnamefont {Calci}},
  \bibinfo {author} {\bibfnamefont {E.}~\bibnamefont {Epelbaum}}, \bibinfo
  {author} {\bibfnamefont {R.}~\bibnamefont {Furnstahl}}, \bibinfo {author}
  {\bibfnamefont {J.}~\bibnamefont {Golak}},  \emph {et~al.} (\bibinfo
  {collaboration} {LENPIC}),\ }\href {\doibase 10.1103/PhysRevC.93.044002}
  {\bibfield  {journal} {\bibinfo  {journal} {Phys. Rev. C}\ }\textbf {\bibinfo
  {volume} {93}},\ \bibinfo {pages} {044002} (\bibinfo {year} {2016})},\
  \Eprint {http://arxiv.org/abs/1505.07218} {arXiv:1505.07218} \BibitemShut
  {NoStop}%
\bibitem [{\citenamefont {Lonardoni}\ \emph {et~al.}(2020)\citenamefont
  {Lonardoni}, \citenamefont {Tews}, \citenamefont {Gandolfi},\ and\
  \citenamefont {Carlson}}]{Lonardoni:2019ypg}%
  \BibitemOpen
  \bibfield  {author} {\bibinfo {author} {\bibfnamefont {D.}~\bibnamefont
  {Lonardoni}}, \bibinfo {author} {\bibfnamefont {I.}~\bibnamefont {Tews}},
  \bibinfo {author} {\bibfnamefont {S.}~\bibnamefont {Gandolfi}}, \ and\
  \bibinfo {author} {\bibfnamefont {J.}~\bibnamefont {Carlson}},\ }\href
  {\doibase 10.1103/PhysRevResearch.2.022033} {\bibfield  {journal} {\bibinfo
  {journal} {Phys.Rev.Res}\ }\textbf {\bibinfo {volume} {2}},\ \bibinfo {pages}
  {022033(R)} (\bibinfo {year} {2020})},\ \Eprint
  {http://arxiv.org/abs/1912.09411} {arXiv:1912.09411} \BibitemShut {NoStop}%
\bibitem [{\citenamefont {{BUQEYE collaboration}}(2020)}]{BUQEYEweb}%
  \BibitemOpen
  \bibfield  {author} {\bibinfo {author} {\bibnamefont {{BUQEYE
  collaboration}}},\ }\href {https://buqeye.github.io/} {} (\bibinfo {year}
  {2020}),\ \bibinfo {note} {\url{https://buqeye.github.io/}}\BibitemShut
  {NoStop}%
\bibitem [{\citenamefont {Melendez}\ \emph {et~al.}(2017)\citenamefont
  {Melendez}, \citenamefont {Wesolowski},\ and\ \citenamefont
  {Furnstahl}}]{Melendez:2017phj}%
  \BibitemOpen
  \bibfield  {author} {\bibinfo {author} {\bibfnamefont {J.~A.}\ \bibnamefont
  {Melendez}}, \bibinfo {author} {\bibfnamefont {S.}~\bibnamefont
  {Wesolowski}}, \ and\ \bibinfo {author} {\bibfnamefont {R.~J.}\ \bibnamefont
  {Furnstahl}},\ }\href {\doibase 10.1103/PhysRevC.96.024003} {\bibfield
  {journal} {\bibinfo  {journal} {Phys. Rev. C}\ }\textbf {\bibinfo {volume}
  {96}},\ \bibinfo {pages} {024003} (\bibinfo {year} {2017})},\ \Eprint
  {http://arxiv.org/abs/1704.03308} {arXiv:1704.03308} \BibitemShut {NoStop}%
\bibitem [{\citenamefont {Drischler}\ \emph
  {et~al.}(2020{\natexlab{a}})\citenamefont {Drischler}, \citenamefont
  {Melendez}, \citenamefont {Furnstahl},\ and\ \citenamefont
  {Phillips}}]{Drischler:2020yad}%
  \BibitemOpen
  \bibfield  {author} {\bibinfo {author} {\bibfnamefont {C.}~\bibnamefont
  {Drischler}}, \bibinfo {author} {\bibfnamefont {J.~A.}\ \bibnamefont
  {Melendez}}, \bibinfo {author} {\bibfnamefont {R.~J.}\ \bibnamefont
  {Furnstahl}}, \ and\ \bibinfo {author} {\bibfnamefont {D.~R.}\ \bibnamefont
  {Phillips}},\ }\href {\doibase 10.1103/PhysRevC.102.054315} {\bibfield
  {journal} {\bibinfo  {journal} {Phys. Rev. C}\ }\textbf {\bibinfo {volume}
  {102}},\ \bibinfo {pages} {054315} (\bibinfo {year} {2020}{\natexlab{a}})},\
  \Eprint {http://arxiv.org/abs/2004.07805} {arXiv:2004.07805 [nucl-th]}
  \BibitemShut {NoStop}%
\bibitem [{\citenamefont {Wesolowski}\ \emph {et~al.}(2019)\citenamefont
  {Wesolowski}, \citenamefont {Furnstahl}, \citenamefont {Melendez},\ and\
  \citenamefont {Phillips}}]{Wesolowski:2018lzj}%
  \BibitemOpen
  \bibfield  {author} {\bibinfo {author} {\bibfnamefont {S.}~\bibnamefont
  {Wesolowski}}, \bibinfo {author} {\bibfnamefont {R.~J.}\ \bibnamefont
  {Furnstahl}}, \bibinfo {author} {\bibfnamefont {J.~A.}\ \bibnamefont
  {Melendez}}, \ and\ \bibinfo {author} {\bibfnamefont {D.~R.}\ \bibnamefont
  {Phillips}},\ }\href {\doibase 10.1088/1361-6471/aaf5fc} {\bibfield
  {journal} {\bibinfo  {journal} {J. Phys. G}\ }\textbf {\bibinfo {volume}
  {46}},\ \bibinfo {pages} {045102} (\bibinfo {year} {2019})},\ \Eprint
  {http://arxiv.org/abs/1808.08211} {arXiv:1808.08211} \BibitemShut {NoStop}%
\bibitem [{\citenamefont {Ireland}\ and\ \citenamefont
  {Nazarewicz}(2015)}]{0954-3899-42-3-030301}%
  \BibitemOpen
  \bibfield  {author} {\bibinfo {author} {\bibfnamefont {D.~G.}\ \bibnamefont
  {Ireland}}\ and\ \bibinfo {author} {\bibfnamefont {W.}~\bibnamefont
  {Nazarewicz}},\ }\href {http://stacks.iop.org/0954-3899/42/i=3/a=030301}
  {\bibfield  {journal} {\bibinfo  {journal} {J. Phys. G Nucl. Part. Phys.}\
  }\textbf {\bibinfo {volume} {42}},\ \bibinfo {pages} {030301} (\bibinfo
  {year} {2015})}\BibitemShut {NoStop}%
\bibitem [{\citenamefont {{{Information and Statistics in Nuclear Experiment
  and Theory (ISNET) Series website}}}(2021)}]{ISNETweb}%
  \BibitemOpen
  \bibfield  {author} {\bibinfo {author} {\bibnamefont {{{Information and
  Statistics in Nuclear Experiment and Theory (ISNET) Series website}}}}\
  }(\bibinfo {year} {2021})\ \bibinfo {note}
  {\url{https://isnet-series.github.io/}}\BibitemShut {NoStop}%
\bibitem [{\citenamefont {{{Bayesian Analysis of Nuclear Dynamics (BAND)
  Framework project}}}(2020)}]{BANDframework}%
  \BibitemOpen
  \bibfield  {author} {\bibinfo {author} {\bibnamefont {{{Bayesian Analysis of
  Nuclear Dynamics (BAND) Framework project}}}}\ }(\bibinfo {year} {2020})\
  \bibinfo {note} {\url{https://bandframework.github.io/}}\BibitemShut
  {NoStop}%
\bibitem [{\citenamefont {Lim}\ and\ \citenamefont {Holt}(2018)}]{lim18}%
  \BibitemOpen
  \bibfield  {author} {\bibinfo {author} {\bibfnamefont {Y.}~\bibnamefont
  {Lim}}\ and\ \bibinfo {author} {\bibfnamefont {J.~W.}\ \bibnamefont {Holt}},\
  }\href {\doibase 10.1103/PhysRevLett.121.062701} {\bibfield  {journal}
  {\bibinfo  {journal} {Phys. Rev. Lett.}\ }\textbf {\bibinfo {volume} {121}},\
  \bibinfo {pages} {062701} (\bibinfo {year} {2018})}\BibitemShut {NoStop}%
\bibitem [{\citenamefont {Carbone}\ \emph {et~al.}(2014)\citenamefont
  {Carbone}, \citenamefont {Rios},\ and\ \citenamefont
  {Polls}}]{Carbone:2014mja}%
  \BibitemOpen
  \bibfield  {author} {\bibinfo {author} {\bibfnamefont {A.}~\bibnamefont
  {Carbone}}, \bibinfo {author} {\bibfnamefont {A.}~\bibnamefont {Rios}}, \
  and\ \bibinfo {author} {\bibfnamefont {A.}~\bibnamefont {Polls}},\ }\href
  {\doibase 10.1103/PhysRevC.90.054322} {\bibfield  {journal} {\bibinfo
  {journal} {Phys. Rev. C}\ }\textbf {\bibinfo {volume} {90}},\ \bibinfo
  {pages} {054322} (\bibinfo {year} {2014})},\ \Eprint
  {http://arxiv.org/abs/1408.0717} {arXiv:1408.0717} \BibitemShut {NoStop}%
\bibitem [{\citenamefont {Drischler}\ \emph
  {et~al.}(2020{\natexlab{b}})\citenamefont {Drischler}, \citenamefont
  {Furnstahl}, \citenamefont {Melendez},\ and\ \citenamefont
  {Phillips}}]{Drischler:2020hwi}%
  \BibitemOpen
  \bibfield  {author} {\bibinfo {author} {\bibfnamefont {C.}~\bibnamefont
  {Drischler}}, \bibinfo {author} {\bibfnamefont {R.~J.}\ \bibnamefont
  {Furnstahl}}, \bibinfo {author} {\bibfnamefont {J.~A.}\ \bibnamefont
  {Melendez}}, \ and\ \bibinfo {author} {\bibfnamefont {D.~R.}\ \bibnamefont
  {Phillips}},\ }\href {\doibase 10.1103/PhysRevLett.125.202702} {\bibfield
  {journal} {\bibinfo  {journal} {Phys. Rev. Lett.}\ }\textbf {\bibinfo
  {volume} {125}},\ \bibinfo {pages} {202702} (\bibinfo {year}
  {2020}{\natexlab{b}})},\ \Eprint {http://arxiv.org/abs/2004.07232}
  {arXiv:2004.07232 [nucl-th]} \BibitemShut {NoStop}%
\bibitem [{\citenamefont {Akmal}\ \emph {et~al.}(1998)\citenamefont {Akmal},
  \citenamefont {Pandharipande},\ and\ \citenamefont {Ravenhall}}]{akmal98}%
  \BibitemOpen
  \bibfield  {author} {\bibinfo {author} {\bibfnamefont {A.}~\bibnamefont
  {Akmal}}, \bibinfo {author} {\bibfnamefont {V.~R.}\ \bibnamefont
  {Pandharipande}}, \ and\ \bibinfo {author} {\bibfnamefont {D.~G.}\
  \bibnamefont {Ravenhall}},\ }\href {\doibase 10.1103/PhysRevC.58.1804}
  {\bibfield  {journal} {\bibinfo  {journal} {Phys. Rev. C}\ }\textbf {\bibinfo
  {volume} {58}},\ \bibinfo {pages} {1804} (\bibinfo {year} {1998})},\ \Eprint
  {http://arxiv.org/abs/nucl-th/9804027} {arXiv:nucl-th/9804027} \BibitemShut
  {NoStop}%
\bibitem [{\citenamefont {Baldo}\ \emph {et~al.}(1997)\citenamefont {Baldo},
  \citenamefont {Bombaci},\ and\ \citenamefont {Burgio}}]{baldo97}%
  \BibitemOpen
  \bibfield  {author} {\bibinfo {author} {\bibfnamefont {M.}~\bibnamefont
  {Baldo}}, \bibinfo {author} {\bibfnamefont {I.}~\bibnamefont {Bombaci}}, \
  and\ \bibinfo {author} {\bibfnamefont {G.~F.}\ \bibnamefont {Burgio}},\
  }\href@noop {} {\bibfield  {journal} {\bibinfo  {journal} {Astron.
  Astrophys.}\ }\textbf {\bibinfo {volume} {328}},\ \bibinfo {pages} {274}
  (\bibinfo {year} {1997})},\ \Eprint {http://arxiv.org/abs/astro-ph/9707277}
  {arXiv:astro-ph/9707277} \BibitemShut {NoStop}%
\bibitem [{\citenamefont {M{\"u}ther}\ \emph {et~al.}(1987)\citenamefont
  {M{\"u}ther}, \citenamefont {Prakash},\ and\ \citenamefont
  {Ainsworth}}]{muether87}%
  \BibitemOpen
  \bibfield  {author} {\bibinfo {author} {\bibfnamefont {H.}~\bibnamefont
  {M{\"u}ther}}, \bibinfo {author} {\bibfnamefont {M.}~\bibnamefont {Prakash}},
  \ and\ \bibinfo {author} {\bibfnamefont {T.~L.}\ \bibnamefont {Ainsworth}},\
  }\href {\doibase https://doi.org/10.1016/0370-2693(87)91611-X} {\bibfield
  {journal} {\bibinfo  {journal} {Phys. Lett. B}\ }\textbf {\bibinfo {volume}
  {199}},\ \bibinfo {pages} {469} (\bibinfo {year} {1987})}\BibitemShut
  {NoStop}%
\bibitem [{\citenamefont {Reinhard}\ \emph {et~al.}(2021)\citenamefont
  {Reinhard}, \citenamefont {Roca-Maza},\ and\ \citenamefont
  {Nazarewicz}}]{Reinhard:2021utv}%
  \BibitemOpen
  \bibfield  {author} {\bibinfo {author} {\bibfnamefont {P.-G.}\ \bibnamefont
  {Reinhard}}, \bibinfo {author} {\bibfnamefont {X.}~\bibnamefont {Roca-Maza}},
  \ and\ \bibinfo {author} {\bibfnamefont {W.}~\bibnamefont {Nazarewicz}},\
  }\href@noop {} {\  (\bibinfo {year} {2021})},\ \Eprint
  {http://arxiv.org/abs/2105.15050} {arXiv:2105.15050} \BibitemShut {NoStop}%
\bibitem [{\citenamefont {H{\"u}ther}\ \emph {et~al.}(2020)\citenamefont
  {H{\"u}ther}, \citenamefont {Vobig}, \citenamefont {Hebeler}, \citenamefont
  {Machleidt},\ and\ \citenamefont {Roth}}]{Huther:2019ont}%
  \BibitemOpen
  \bibfield  {author} {\bibinfo {author} {\bibfnamefont {T.}~\bibnamefont
  {H{\"u}ther}}, \bibinfo {author} {\bibfnamefont {K.}~\bibnamefont {Vobig}},
  \bibinfo {author} {\bibfnamefont {K.}~\bibnamefont {Hebeler}}, \bibinfo
  {author} {\bibfnamefont {R.}~\bibnamefont {Machleidt}}, \ and\ \bibinfo
  {author} {\bibfnamefont {R.}~\bibnamefont {Roth}},\ }\href {\doibase
  10.1016/j.physletb.2020.135651} {\bibfield  {journal} {\bibinfo  {journal}
  {Phys. Lett. B}\ }\textbf {\bibinfo {volume} {808}},\ \bibinfo {pages}
  {135651} (\bibinfo {year} {2020})},\ \Eprint
  {http://arxiv.org/abs/1911.04955} {arXiv:1911.04955} \BibitemShut {NoStop}%
\bibitem [{\citenamefont {Drischler}\ \emph
  {et~al.}(2021{\natexlab{c}})\citenamefont {Drischler}, \citenamefont {Han},
  \citenamefont {Lattimer}, \citenamefont {Prakash}, \citenamefont {Reddy},\
  and\ \citenamefont {Zhao}}]{Drischler:2020fvz}%
  \BibitemOpen
  \bibfield  {author} {\bibinfo {author} {\bibfnamefont {C.}~\bibnamefont
  {Drischler}}, \bibinfo {author} {\bibfnamefont {S.}~\bibnamefont {Han}},
  \bibinfo {author} {\bibfnamefont {J.~M.}\ \bibnamefont {Lattimer}}, \bibinfo
  {author} {\bibfnamefont {M.}~\bibnamefont {Prakash}}, \bibinfo {author}
  {\bibfnamefont {S.}~\bibnamefont {Reddy}}, \ and\ \bibinfo {author}
  {\bibfnamefont {T.}~\bibnamefont {Zhao}},\ }\href {\doibase
  10.1103/PhysRevC.103.045808} {\bibfield  {journal} {\bibinfo  {journal}
  {Phys. Rev. C}\ }\textbf {\bibinfo {volume} {103}},\ \bibinfo {pages}
  {045808} (\bibinfo {year} {2021}{\natexlab{c}})},\ \Eprint
  {http://arxiv.org/abs/2009.06441} {arXiv:2009.06441 [nucl-th]} \BibitemShut
  {NoStop}%
\bibitem [{\citenamefont {Drischler}\ \emph
  {et~al.}(2021{\natexlab{d}})\citenamefont {Drischler}, \citenamefont
  {Quinonez}, \citenamefont {Giuliani}, \citenamefont {Lovell},\ and\
  \citenamefont {Nunes}}]{Drischler:2021qoy}%
  \BibitemOpen
  \bibfield  {author} {\bibinfo {author} {\bibfnamefont {C.}~\bibnamefont
  {Drischler}}, \bibinfo {author} {\bibfnamefont {M.}~\bibnamefont {Quinonez}},
  \bibinfo {author} {\bibfnamefont {P.~G.}\ \bibnamefont {Giuliani}}, \bibinfo
  {author} {\bibfnamefont {A.~E.}\ \bibnamefont {Lovell}}, \ and\ \bibinfo
  {author} {\bibfnamefont {F.~M.}\ \bibnamefont {Nunes}},\ }\href@noop {} {\
  (\bibinfo {year} {2021}{\natexlab{d}})},\ \Eprint
  {http://arxiv.org/abs/2108.08269} {arXiv:2108.08269} \BibitemShut {NoStop}%
\bibitem [{\citenamefont {Zhang}\ and\ \citenamefont
  {Furnstahl}()}]{Zhang:2021xx}%
  \BibitemOpen
  \bibfield  {author} {\bibinfo {author} {\bibfnamefont {X.}~\bibnamefont
  {Zhang}}\ and\ \bibinfo {author} {\bibfnamefont {R.~J.}\ \bibnamefont
  {Furnstahl}},\ }\href@noop {} {\ }\bibinfo {note} {{in
  preparation}}\BibitemShut {NoStop}%
\bibitem [{\citenamefont {Abbott}\ \emph {et~al.}(2018)\citenamefont {Abbott}
  \emph {et~al.}}]{Abbott:2018exr}%
  \BibitemOpen
  \bibfield  {author} {\bibinfo {author} {\bibfnamefont {B.}~\bibnamefont
  {Abbott}} \emph {et~al.} (\bibinfo {collaboration} {LIGO Scientific,
  Virgo}),\ }\href {\doibase 10.1103/PhysRevLett.121.161101} {\bibfield
  {journal} {\bibinfo  {journal} {Phys. Rev. Lett.}\ }\textbf {\bibinfo
  {volume} {121}},\ \bibinfo {pages} {161101} (\bibinfo {year} {2018})},\
  \Eprint {http://arxiv.org/abs/1805.11581} {arXiv:1805.11581} \BibitemShut
  {NoStop}%
\bibitem [{\citenamefont {De}\ \emph {et~al.}(2018)\citenamefont {De},
  \citenamefont {Finstad}, \citenamefont {Lattimer}, \citenamefont {Brown},
  \citenamefont {Berger},\ and\ \citenamefont {Biwer}}]{De:2018uhw}%
  \BibitemOpen
  \bibfield  {author} {\bibinfo {author} {\bibfnamefont {S.}~\bibnamefont
  {De}}, \bibinfo {author} {\bibfnamefont {D.}~\bibnamefont {Finstad}},
  \bibinfo {author} {\bibfnamefont {J.~M.}\ \bibnamefont {Lattimer}}, \bibinfo
  {author} {\bibfnamefont {D.~A.}\ \bibnamefont {Brown}}, \bibinfo {author}
  {\bibfnamefont {E.}~\bibnamefont {Berger}}, \ and\ \bibinfo {author}
  {\bibfnamefont {C.~M.}\ \bibnamefont {Biwer}},\ }\href {\doibase
  10.1103/PhysRevLett.121.091102} {\bibfield  {journal} {\bibinfo  {journal}
  {Phys. Rev. Lett.}\ }\textbf {\bibinfo {volume} {121}},\ \bibinfo {pages}
  {091102} (\bibinfo {year} {2018})},\ \bibinfo {note} {[Erratum: Phys. Rev.
  Lett. \textbf{121}, 259902 (2018)]},\ \Eprint
  {http://arxiv.org/abs/1804.08583} {arXiv:1804.08583} \BibitemShut {NoStop}%
\bibitem [{\citenamefont {Capano}\ \emph {et~al.}(2020)\citenamefont {Capano},
  \citenamefont {Tews}, \citenamefont {Brown}, \citenamefont {Margalit},
  \citenamefont {De}, \citenamefont {Kumar}, \citenamefont {Brown},
  \citenamefont {Krishnan},\ and\ \citenamefont {Reddy}}]{Capano:2019eae}%
  \BibitemOpen
  \bibfield  {author} {\bibinfo {author} {\bibfnamefont {C.~D.}\ \bibnamefont
  {Capano}}, \bibinfo {author} {\bibfnamefont {I.}~\bibnamefont {Tews}},
  \bibinfo {author} {\bibfnamefont {S.~M.}\ \bibnamefont {Brown}}, \bibinfo
  {author} {\bibfnamefont {B.}~\bibnamefont {Margalit}}, \bibinfo {author}
  {\bibfnamefont {S.}~\bibnamefont {De}}, \bibinfo {author} {\bibfnamefont
  {S.}~\bibnamefont {Kumar}}, \bibinfo {author} {\bibfnamefont {D.~A.}\
  \bibnamefont {Brown}}, \bibinfo {author} {\bibfnamefont {B.}~\bibnamefont
  {Krishnan}}, \ and\ \bibinfo {author} {\bibfnamefont {S.}~\bibnamefont
  {Reddy}},\ }\href {\doibase 10.1038/s41550-020-1014-6} {\bibfield  {journal}
  {\bibinfo  {journal} {Nature Astron.}\ }\textbf {\bibinfo {volume} {4}},\
  \bibinfo {pages} {625} (\bibinfo {year} {2020})},\ \Eprint
  {http://arxiv.org/abs/1908.10352} {arXiv:1908.10352} \BibitemShut {NoStop}%
\bibitem [{\citenamefont {Miller}\ \emph
  {et~al.}(2021{\natexlab{b}})\citenamefont {Miller} \emph
  {et~al.}}]{Miller:2021qha}%
  \BibitemOpen
  \bibfield  {author} {\bibinfo {author} {\bibfnamefont {M.~C.}\ \bibnamefont
  {Miller}} \emph {et~al.},\ }\href@noop {} {\  (\bibinfo {year}
  {2021}{\natexlab{b}})},\ \Eprint {http://arxiv.org/abs/2105.06979}
  {arXiv:2105.06979} \BibitemShut {NoStop}%
\bibitem [{\citenamefont {Riley}\ \emph {et~al.}(2021)\citenamefont {Riley}
  \emph {et~al.}}]{Riley:2021pdl}%
  \BibitemOpen
  \bibfield  {author} {\bibinfo {author} {\bibfnamefont {T.~E.}\ \bibnamefont
  {Riley}} \emph {et~al.},\ }\href@noop {} {\  (\bibinfo {year} {2021})},\
  \Eprint {http://arxiv.org/abs/2105.06980} {arXiv:2105.06980} \BibitemShut
  {NoStop}%
\bibitem [{\citenamefont {Adhikari}\ \emph {et~al.}(2021)\citenamefont
  {Adhikari} \emph {et~al.}}]{PREX:2021umo}%
  \BibitemOpen
  \bibfield  {author} {\bibinfo {author} {\bibfnamefont {D.}~\bibnamefont
  {Adhikari}} \emph {et~al.} (\bibinfo {collaboration} {PREX}),\ }\href
  {\doibase 10.1103/PhysRevLett.126.172502} {\bibfield  {journal} {\bibinfo
  {journal} {Phys. Rev. Lett.}\ }\textbf {\bibinfo {volume} {126}},\ \bibinfo
  {pages} {172502} (\bibinfo {year} {2021})},\ \Eprint
  {http://arxiv.org/abs/2102.10767} {arXiv:2102.10767} \BibitemShut {NoStop}%
\bibitem [{\citenamefont {Reed}\ \emph {et~al.}(2021)\citenamefont {Reed},
  \citenamefont {Fattoyev}, \citenamefont {Horowitz},\ and\ \citenamefont
  {Piekarewicz}}]{Reed:2021nqk}%
  \BibitemOpen
  \bibfield  {author} {\bibinfo {author} {\bibfnamefont {B.~T.}\ \bibnamefont
  {Reed}}, \bibinfo {author} {\bibfnamefont {F.~J.}\ \bibnamefont {Fattoyev}},
  \bibinfo {author} {\bibfnamefont {C.~J.}\ \bibnamefont {Horowitz}}, \ and\
  \bibinfo {author} {\bibfnamefont {J.}~\bibnamefont {Piekarewicz}},\ }\href
  {\doibase 10.1103/PhysRevLett.126.172503} {\bibfield  {journal} {\bibinfo
  {journal} {Phys. Rev. Lett.}\ }\textbf {\bibinfo {volume} {126}},\ \bibinfo
  {pages} {172503} (\bibinfo {year} {2021})},\ \Eprint
  {http://arxiv.org/abs/2101.03193} {arXiv:2101.03193} \BibitemShut {NoStop}%
\bibitem [{\citenamefont {Furnstahl}\ \emph {et~al.}(2021)\citenamefont
  {Furnstahl}, \citenamefont {Hammer},\ and\ \citenamefont
  {Schwenk}}]{Furnstahl:2021rfk}%
  \BibitemOpen
  \bibfield  {author} {\bibinfo {author} {\bibfnamefont {R.~J.}\ \bibnamefont
  {Furnstahl}}, \bibinfo {author} {\bibfnamefont {H.~W.}\ \bibnamefont
  {Hammer}}, \ and\ \bibinfo {author} {\bibfnamefont {A.}~\bibnamefont
  {Schwenk}},\ }\href {\doibase 10.1007/s00601-021-01658-5} {\bibfield
  {journal} {\bibinfo  {journal} {Few Body Syst.}\ }\textbf {\bibinfo {volume}
  {62}},\ \bibinfo {pages} {72} (\bibinfo {year} {2021})},\ \Eprint
  {http://arxiv.org/abs/2107.00413} {arXiv:2107.00413 [nucl-th]} \BibitemShut
  {NoStop}%
\end{thebibliography}%

\end{document}